\documentclass[preprint]{aastex} 
\usepackage{epsfig}
\usepackage{natbib}
\usepackage{graphicx}
\usepackage{graphics}
\usepackage{subfig}

\shorttitle{V1647 Ori Optical Outburst X-ray Production}
\shortauthors{Teets et al.}

\begin{document}

\title{X-ray Production by V1647 Ori During Optical Outbursts}

\author{William K. Teets\altaffilmark{1}, David A. Weintraub\altaffilmark{1}, Nicolas Grosso\altaffilmark{2}, David Principe\altaffilmark{4},\\
 Joel H. Kastner\altaffilmark{4}, Kenji Hamaguchi\altaffilmark{3}, Michael Richmond\altaffilmark{4}}

\altaffiltext{1}{Department of Physics \& Astronomy, Vanderbilt University, Nashville, TN 37235, USA}
\altaffiltext{2}{Observatoire Astronomique de Strasbourg, Universit\'e de Strasbourg, CNRS, UMR 7550, 11 rue de l'Universit\'e, F-67000 Strasbourg, France}
\altaffiltext{3}{Goddard Space Flight Center, Greenbelt, MD 20771, USA}
\altaffiltext{4}{Rochester Institute of Technology, Rochester, NY 14623-5604, USA}

\begin{abstract}

The pre-main sequence star V1647 Ori has recently undergone two optical/near-infrared (OIR) outbursts that are associated with dramatic enhancements in the stellar accretion rate. Our intensive X-ray monitoring of this object affords the opportunity to investigate whether and how the intense X-ray emission is related to pre-MS accretion activity. Our analysis of all fourteen Chandra X-ray Observatory observations of V1647 Ori demonstrate that variations in the X-ray luminosity of V1647 Ori are correlated with similar changes in the OIR brightness of this source during both (2003--2005 and 2008) eruptions, strongly supporting the hypothesis that accretion is the primary generation mechanism for the X-ray outbursts. Furthermore, the Chandra monitoring demonstrates that the X-ray spectral properties of the second eruption were strikingly similar to those of the 2003 eruption. We find that X-ray spectra obtained immediately following the second outburst --- during which V1647 Ori exhibited high X-ray luminosities, high hardness ratios, and strong X-ray variability --- are well modeled as a heavily absorbed (N$_{H}$ $\sim$ 4 $\times$10$^{22}$ cm$^{-2}$), single-component plasma with characteristic temperatures (kT$_{X}$ $\sim$ 2--6 keV) that are consistently too high to be generated via accretion shocks but are in the range expected for plasma heated by magnetic reconnection events. We also find that the X-ray absorbing column has not changed significantly throughout the observing campaign. Since the OIR and X-ray changes are correlated, we hypothesize that these reconnection events either occur in the accretion stream connecting the circumstellar disk to the star or in accretion-enhanced protostellar coronal activity.

\end{abstract}

\section{Introduction}\label{intro}
The young stellar object (YSO) \object{V1647 Ori} was first noted to be an erupting source when it brightened suddenly in November 2003, illuminating a new nebula now known as McNeil's Nebula.  This deeply embedded, low-mass YSO is typically considered to be an EX Lupi-type object (an ``EXor") though it shares some spectral characteristics with FU Orionis \citep{vac04}, the prototype of a similar class of erupting pre-main sequence (PMS) stars (``FUors").  EXors are observed to brighten irregularly at optical wavelengths up to several times per decade; these outbursts persist for weeks to months \citep{her01}.  FUors are YSOs that erupt less often than EXors, perhaps only once per century, and fade much more slowly, i.e., on timescales of years to decades \citep{her77}.  
The general consensus is that these eruptions are the result of massive accretion events that occur irregularly as young protostars grow; such accretion episodes may be the primary process through which young stars accrete most of their mass \citep{har96}. The mechanisms underlying EXor and FUor outbursts remain very poorly understood; however, models generally invoke the rapid onset of disk instabilities that lead to sudden inward migration of the inner disk truncation radius and dramatic changes in the star/disk magnetic  field configuration (e.g., \citet{zhu09, kon11}).

Soft X-ray ($\sim$0.3 keV) generation via shock-heating of plasma at the bases of PMS accretion streams has been proposed for some objects from X-ray grating observations \citep[see review by][]{gud09} using plasma diagnostics based on helium-like ions \citep[see review by][] {por10}, a prime example being TW Hydrae \citep{kas02,brick10}.  On the other hand, studies such as the Chandra Orion Ultradeep Project (COUP) indicate that coronal activity, and not accretion, is likely the primary production mechanism for the harder X-rays that are characteristic of most pre-main sequence stars \citep{pre05,sta06}.  Debate still lingers as to the primary X-ray generation mechanism in YSOs during optical outbursts.    

Since 2003, V1647 Ori has been observed to undergo two distinct outbursts that have been detected in the optical and near-infrared \citep{mcn03,ita08} as well as at X-ray wavelengths \citep{kas04,kas06,gro05, gro06,ham10}.  The magnitudes and timescales for brightening and fading displayed by V1647 Ori do not match those of EXors.  Furthermore, the duration of the initial outburst was shorter than those of FUors  \citep{kas06,asp08}.  However, spectroscopic data do show signatures of accretion, such as strong H$\alpha$ and Br$\gamma$ emission \citep{rei04,bri10}, confirming that the eruptions of V1647 Ori do resemble those of FUors and EXors. 

V1647 Ori has been observed in X-rays before and during both of the eruptions (as well as after the first of the two eruptions) detected in the optical/near-infrared since 2003.  Although other FUor- or EXor-like YSOs have been observed in X-rays (V1118 Ori \citep{aud10}; EX Lup \citep{gro10}; Z CMa \citep{ste09}; FU Ori \citep{ski06}; V1735 Cyg \citep{ski09}), V1647 Ori is the only eruptive YSO to undergo such extensive X-ray monitoring.  During the 2003 eruption, the sudden rise in flux and subsequent decline in both the optical and near-infrared correlated strongly with a sharp increase and then decline in the X-ray flux, which suggests a common origin \citep{kas04,kas06}.  In this paper, we present and analyze X-ray spectra obtained with the Chandra X-ray Observatory (CXO) during the 2008--2009 outburst.  In \S\ref{obs}, we describe the observations and data reduction.  In \S\ref{results}, we examine the trends and patterns seen in the X-ray emission from V1647 Ori over the past two outbursts, present modeling results for each of the five recent CXO observations, and compare these results to those obtained from previous (2003--2005) CXO and XMM-Newton observations. Finally, in \S\ref{discussion} we discuss the implications of the results, and we argue that the measured properties of the X-ray emitting plasma are best understood within the framework of the accretion-generated X-ray hypothesis.  

\section{Observations \& Data Reduction}\label{obs}
Observations of V1647 Ori made with the Chandra X-ray Observatory (CXO) were triggered in 2008 September with a 20 ks observation (CXO Cycle 10, PI: D. Weintraub, ObsID 9915), after V1647 Ori was reported to have undergone a new optical outburst between early 2008 January and late 2008 August \citep{ita08}.  Subsequent 20 ks observations were initiated in 2009 January and April (ObsIDs 9916 and 9917, PI: D. Weintraub).  In addition, in 2008 November, two observations (ObsIDs 10763 and 8585; PI: N. Calvet) were made of the NGC 2068/2071 region that serendipitously included V1647 Ori.  Together, these five CXO pointings yield an extended sequence of observations of this source over a seven-month period immediately following the onset of the optical outburst in 2008.

Fortuitously, a field of view that includes V1647 Ori was observed in 2002 November, a full year before the start of the 2003 outburst \citep{sim04}; subsequent CXO observations targeting V1647 Ori were obtained in 2004 March and April, 2005 August and December, and 2006 May and August \citep{kas04,kas06}.  In addition, targeted XMM-Newton observations were obtained in 2004 April \citep{gro05} and 2005 March \citep{gro06}, and a targeted Suzaku observation was obtained in 2008 October \citep{ham10}.  The 2004--2006 observations cover a two-year period during which V1647 Ori more or less faded steadily from a strong YSO X-ray source into a very faint one, albeit with large X-ray flux variability.  For direct comparison with our analysis of the 2008--2009 epoch datasets, we have re-reduced and analyzed all of the 2002--2006 CXO datasets, using the same techniques and software packages that we have used for the latest (2008--2009) datasets.

For all Chandra observations, the Advanced CCD Imaging Spectrometer (ACIS) was used in one of two imaging configurations. ACIS detectors have a pixel size of 0.49$\arcsec$ and each ACIS-I and ACIS-S CCD has a field of view of 8.3$\arcmin$ $\times$ 8.3$\arcmin$.  CXO/ACIS has significant sensitivity over the energy range 0.3--10 keV, with the soft ($<$1 keV) X-ray sensitivity dependent on whether a back-illuminated or front-illuminated CCD serves as the detector (see below).  The CIAO v4.1 software package and CALDB v4.1.0--4.1.4 calibration files were used to reduce the data and extract pulse-invariant (PI) spectra.  Observation details are given in Table~\ref{tab-1}.  To compare X-ray luminosities from 2008--2009 with those obtained from previous observations, source spectra were extracted from 2.5$\arcsec$ radius regions (making sure the aperture was positioned so as to encompass as many source photons as possible, even when V1647 Ori is 4$\arcmin$ off-axis) while background spectra were extracted from regions near but beyond 2.5$\arcsec$ from the target, on the same chip, using 20$\arcsec$ outer radius extraction apertures.  Inspection of background light curves reveals background levels to be fairly constant with no evidence of large fluctuations, such as flares, occurring during the observations.  Spectra were also re-extracted from previous Chandra observations using the same apertures to ensure that direct comparisons of results from different observation epochs could be performed reliably. For observations in which the primary target was not V1647 Ori (ObsIDs 2539, 10763, and 8585), exposure maps were generated to investigate off-axis position effects.  In all three of these observations, the net count and mean count rate corrections due to off-axis source positions were insignificant, on the order of a few percent.  The resultant spectral data points of the five recent observations were grouped into energy bins with a minimum of five counts per bin before spectral modeling was done.  The count rates were high enough and durations long enough for each of the 2008--2009 observations that this bin size yielded PI spectra with good statistics.  For the 2002--2006 observations, we employed single-count-minimum binning for very low-count observations and five-count-minimum binning for higher-count observations so as to yield spectra suitable for modeling.  Because no events with energies less than 0.5 keV and few events with energies greater than 8.0 keV were detected from V1647 Ori, we limited our modeling to the 0.5--8.0 keV energy range. 

Since V1647 Ori was imaged with both front- and back-illuminated CCDs in the exposure sequence under analysis, we generated synthetic spectra and convolved these spectra with the instrument responses to determine whether mean, broadband (0.5--8.0 keV) count rates were directly comparable for all observations of V1647 Ori.  The simulations (Table~\ref{tab-fibi}) showed that, for the plasma temperature regimes considered here, the back-illuminated S3 CCD is $\sim$10\% more sensitive to incoming flux than the front-illuminated ACIS-I CCDs.  Net counts and mean count rates in the energy range 0.5--8.0 keV (Table~\ref{tab-1}) were therefore adjusted downward accordingly for the two observations (ObsIDs 5307 and 5308) that used the S3 CCD in order to remove this sensitivity bias.

\section{Results}\label{results} 

\subsection{Short-Term and Long-Term Variability}\label{var}
 
\begin{figure}
\epsscale{1.0}
\plotone{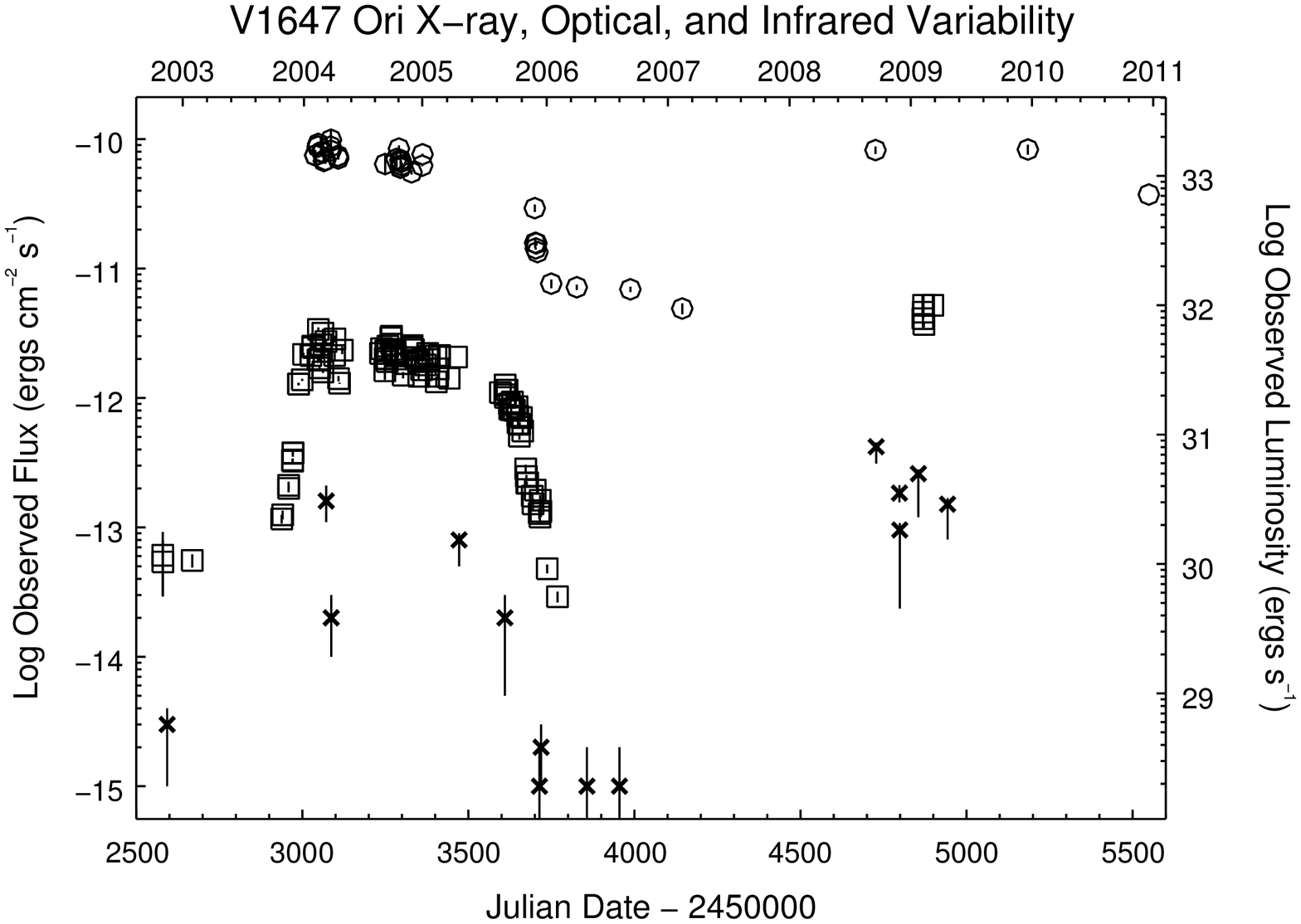}
\caption{Near-infrared and X-ray light curves of V1647 Ori.  Chandra X-ray data are shown as crosses; the I$_{c}$-band data are shown as squares, and the H-band data are shown as circles.  I-band Goddard Robotic Telescope (GRT) \citep{sak11} data (Michael Richmond, personal communication) were obtained in 2009. For the published I- and H-band data, error bar size is on the order of the data point size (data are from:  \citet{bri04}, \citet{mcg04}, \citet{rei04}, \citet{sem04}, \citet{ojh05}, \citet{ojh06}, \citet{sem06}, \citet{ven06}, \citet{aco07}, \citet{fed07a}, \citet{asp08}, \citet{ojh08}, \citet{asp09b}, and \citet{ven11}).  The uncertainties for the GRT data and CXO data are 1$\sigma$.  Calendar year is indicated along the top horizontal axis.  \label{overall}}
\end{figure}

We calculated the median photon energies, mean count rates, and mean hardness ratios for all 14 CXO observations of V1647 Ori.  These results are reported in Table~\ref{tab-1}.  In calculating the hardness ratios, the hard (H) X-rays are defined as those in the range from 2.8 to 8.0 keV and the soft (S) X-rays as those with energies from 0.5 to 2.8 keV \citep{gro05}.  The hardness ratio is then HR = (H-S)/(H+S), such that negative values of HR indicate softer spectra, and positive values indicate harder spectra.  The recalculated values and error ranges of the mean count rates and median photon energies for the 2004--2006 observations differ only marginally from those reported by \citet{kas06}.  With data from only three epochs in hand, \citet{kas04} reported that the X-ray flux brightened and hardened during the outburst (2002 November 14 to 2004 March 07) and then quickly faded and softened post-outburst (2004 March 07 to 2004 March 22).  With data from additional epochs available, however, \citet{kas06} reinterpreted the late 2004 March data as a short-term downward fluctuation, not as a quick end to the original outburst.  They interpret the observed pattern of changes in the median photon energy as evidence that the X-ray spectrum hardened during outburst, remained somewhat hard for at least one year, and then softened after 2005 August 27 as both the optical and X-ray flux from V1647 Ori returned to pre-outburst levels. The results obtained here (for hardness ratio) confirm this general trend.  This hardening and softening (the mean hardness ratios (with 1$\sigma$ uncertainties) changed from $-$0.34 $\pm$ 0.31 (soft) to +0.46 $\pm$ 0.15 (hard) and back to $-$0.45 $\pm$ 0.33 (very soft) over the course of three years, from 2002 November through 2005 December) took place over a timescale of months, with V1647 Ori remaining near peak outburst X-ray luminosity and hardness for more than one year - with the notable exception of the 2004 March 22 data.  V1647 Ori likely remained in its soft X-ray, low X-ray luminosity phase for almost three years, from late 2005 until its latest optical outburst in 2008 August.  The five CXO observations covering the interval from 2008 September through 2009 April all show V1647 Ori in a hard X-ray, high X-ray luminosity state.  As was the case during the 2002--2006 observations, the source remained near peak X-ray luminosity for at least seven months (Fig.~\ref{overall}, figure to be discussed in Section \ref{discussion}).  As of 2009 December 18, V1647 Ori appeared to still be in an outburst phase with J-, H-, and K-band photometric magnitudes near values measured at the brightest phase of outburst during 2008 September \citep{ven09}. Thus, it appears that V1647 Ori remained near peak luminosity for at least one year.\footnotemark[1]

\footnotetext[1]{As of submission of this paper, the outburst appears to be ongoing.}

\begin{figure}
\centering
\begin{tabular}{ccc}
\plotfiddle{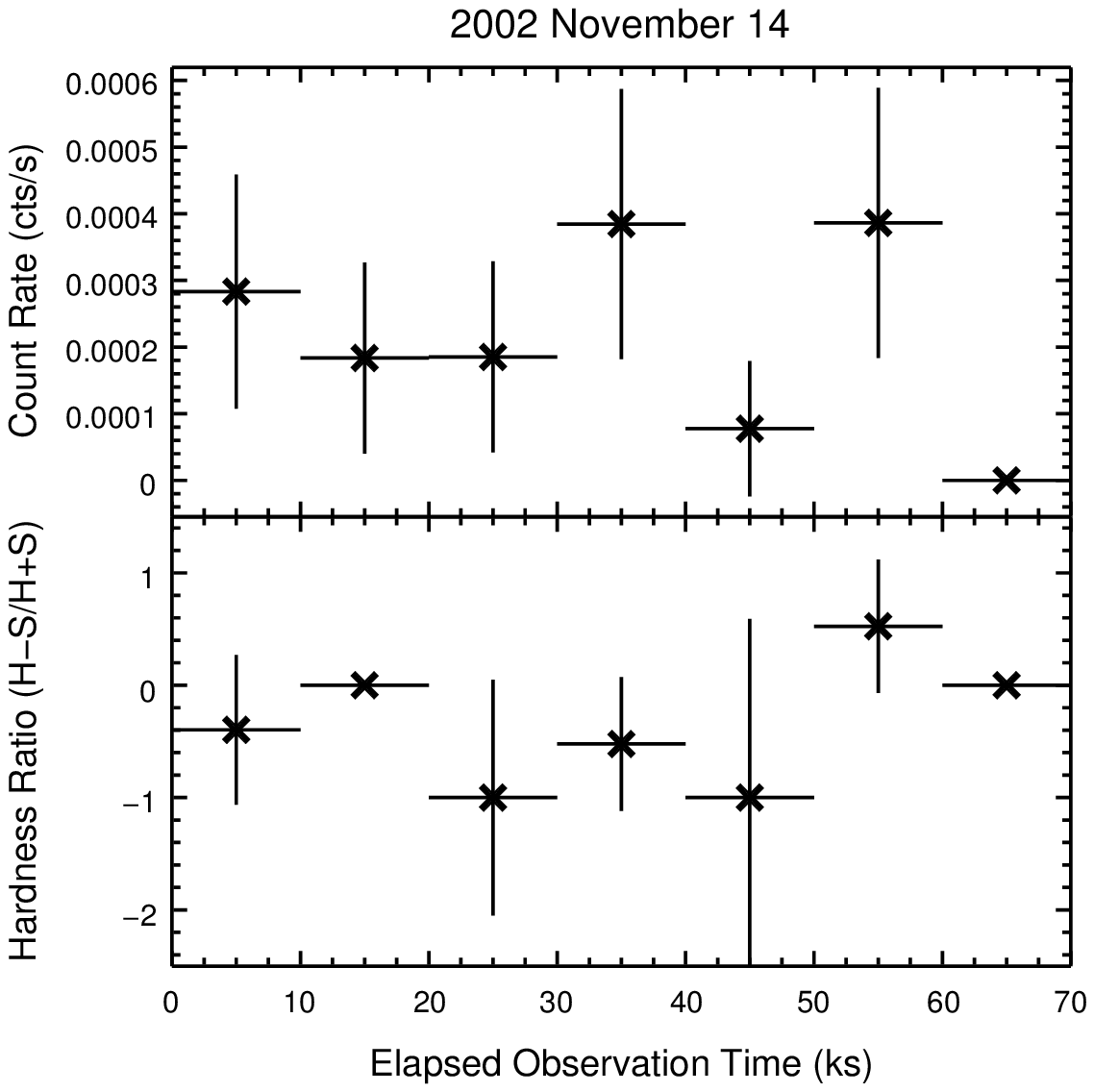}{3.0cm}{0}{170}{170}{-525}{-100} & \plotfiddle{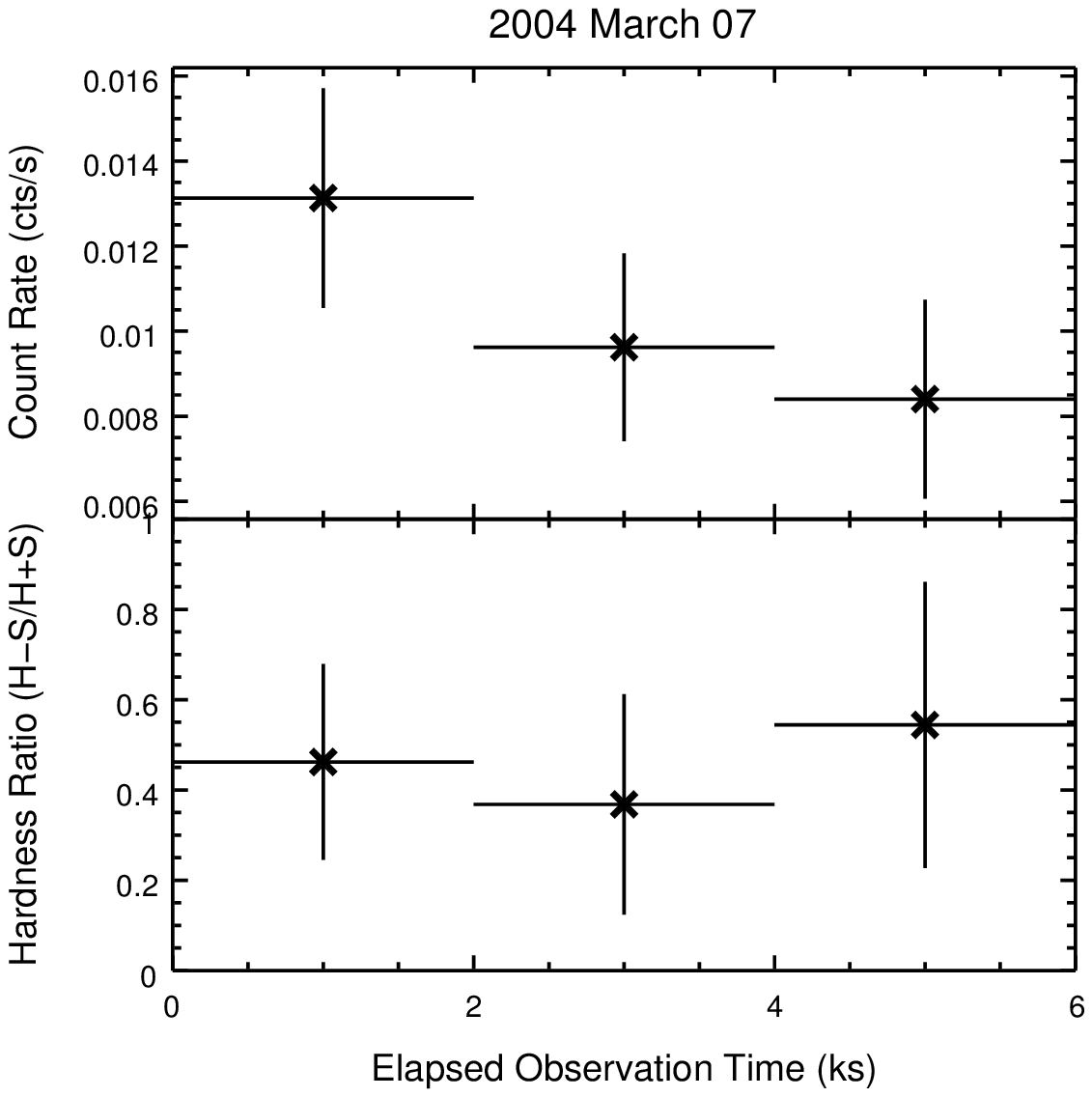}{3.0cm}{0}{170}{170}{-475}{-100} & \plotfiddle{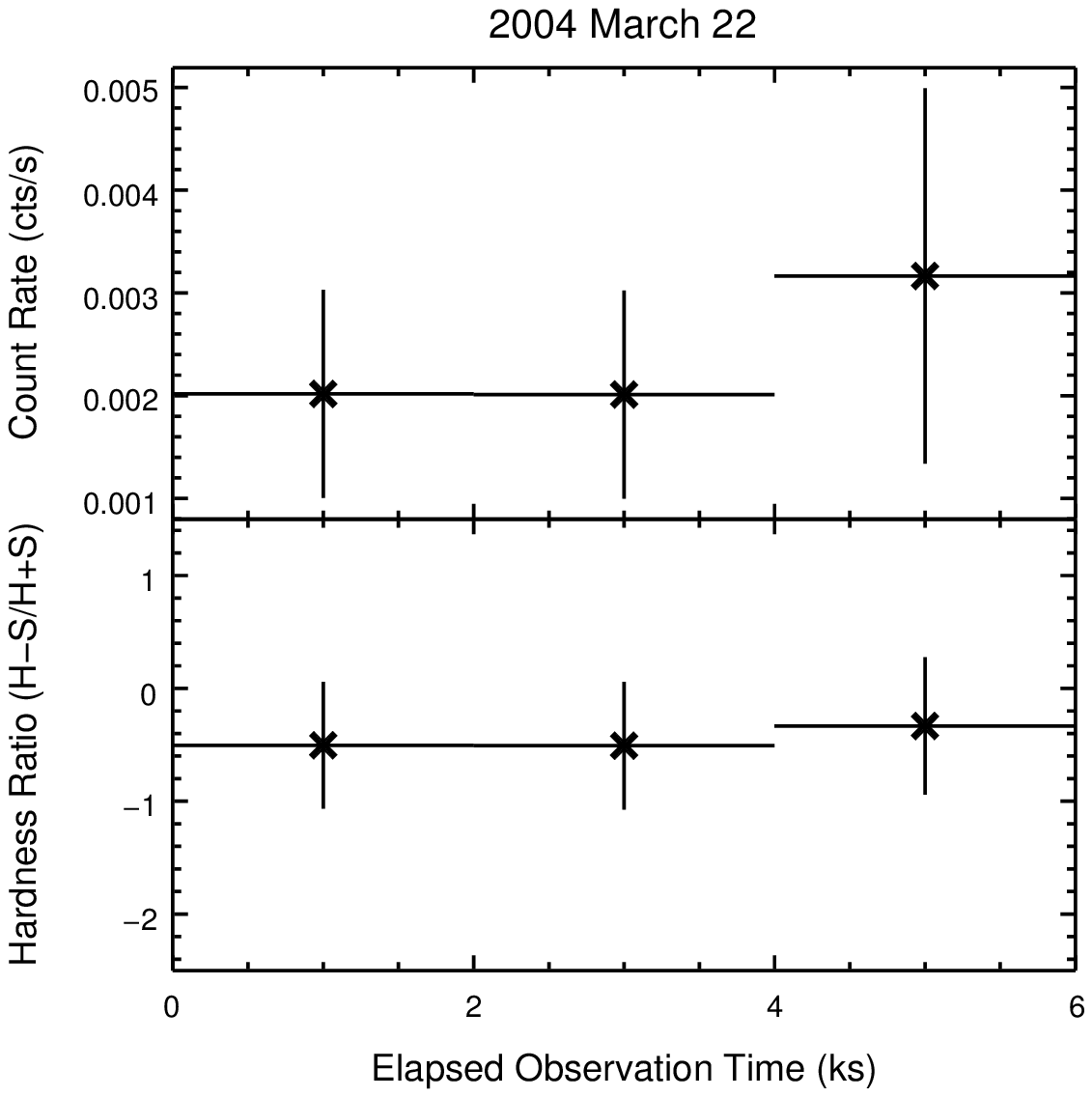}{3.0cm}{0}{170}{170}{-475}{-100} \\\\
\plotfiddle{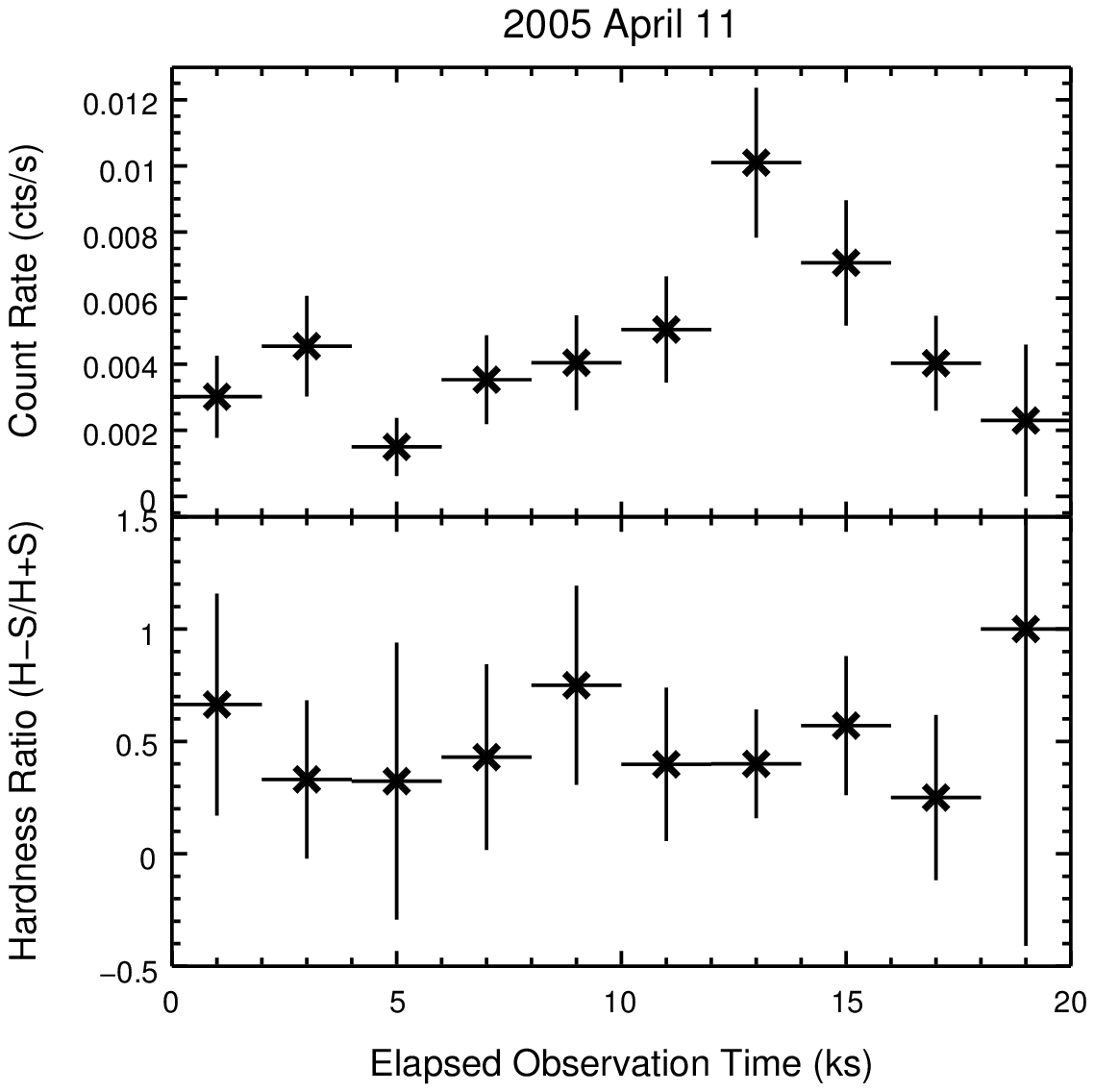}{3.0cm}{0}{170}{170}{-525}{-100} & \plotfiddle{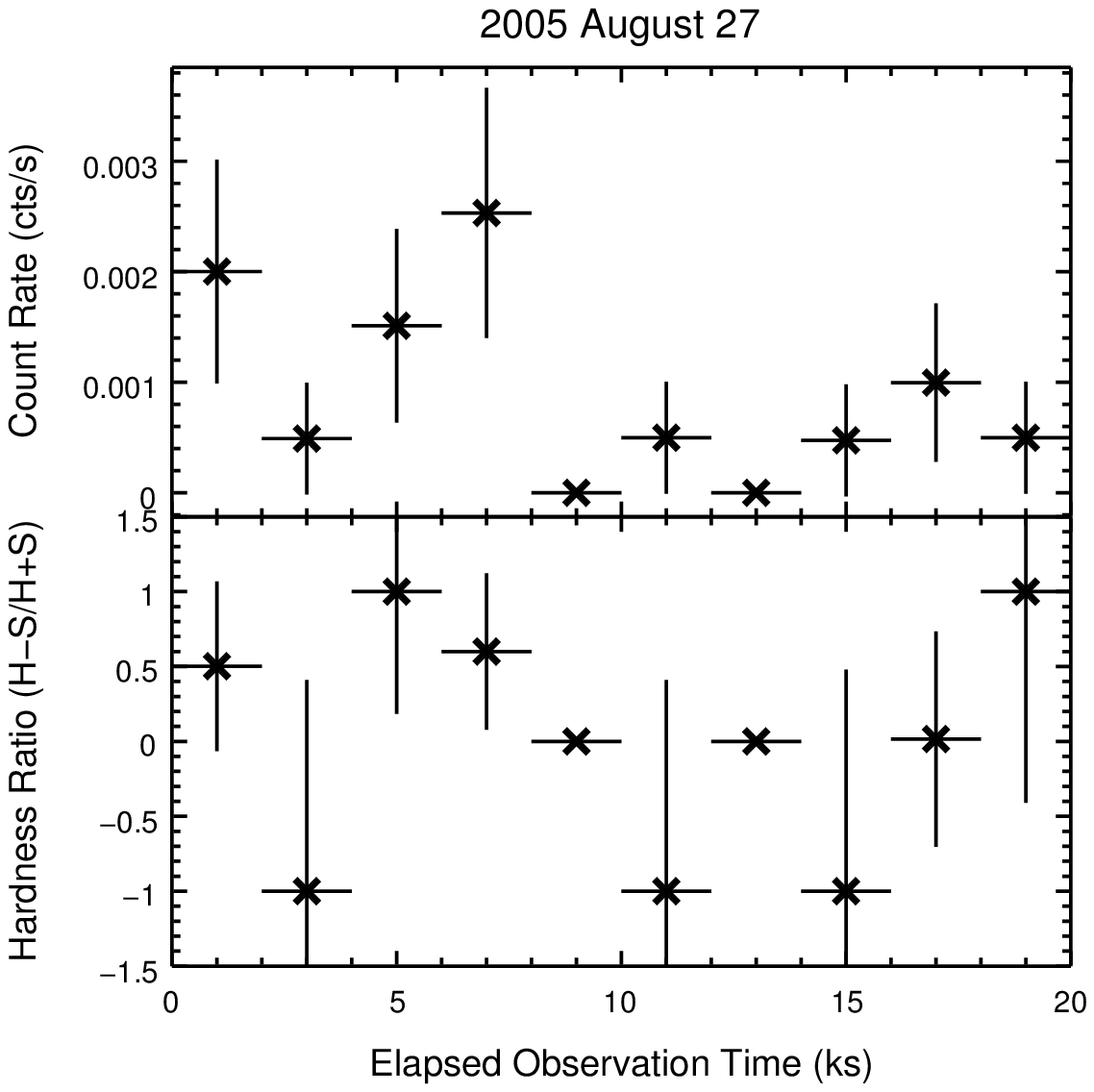}{3.0cm}{0}{170}{170}{-475}{-100} & \plotfiddle{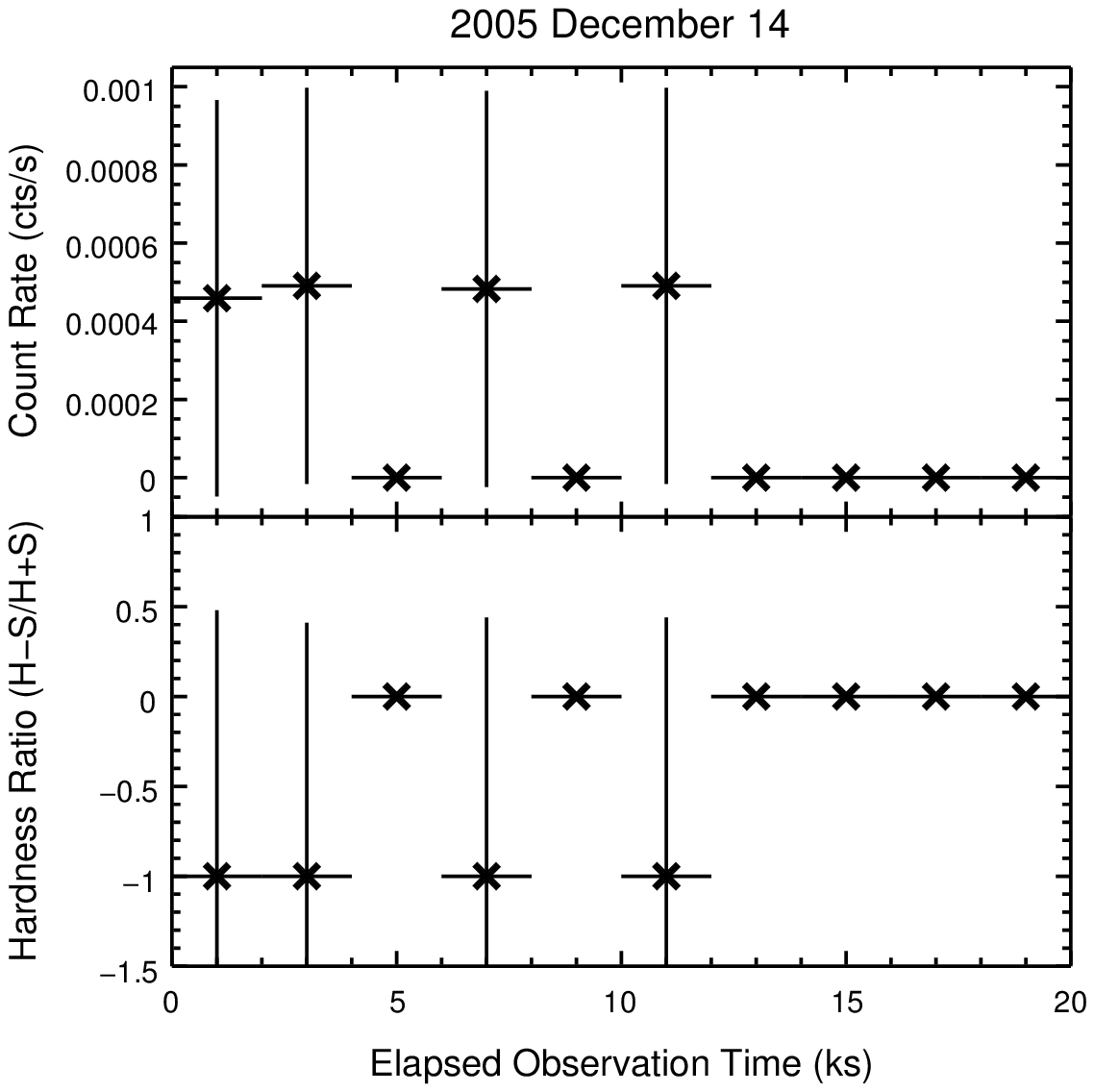}{3.0cm}{0}{170}{170}{-475}{-100} \\ 
\end{tabular}
\caption{Background-subtracted X-ray light curves for 2002-2006 epoch observations.  Time bins are 2 ks for each observation (except 2002 November 14, which uses 10 ks bins due to the very low count rate) and contain counts (0.5--8.0 keV) per total time associated with each bin, not the total time associated with the observation. 
Observations for 2005 December 09, 2006 May 01, and 2006 August 07 do not detect V1647 Ori, so their light curves are not presented.  Uncertainties in mean hardness ratios and count rates are 1$\sigma$.\label{lcprev}}
\end{figure}

\begin{figure}
\epsscale{0.75}
\plotone{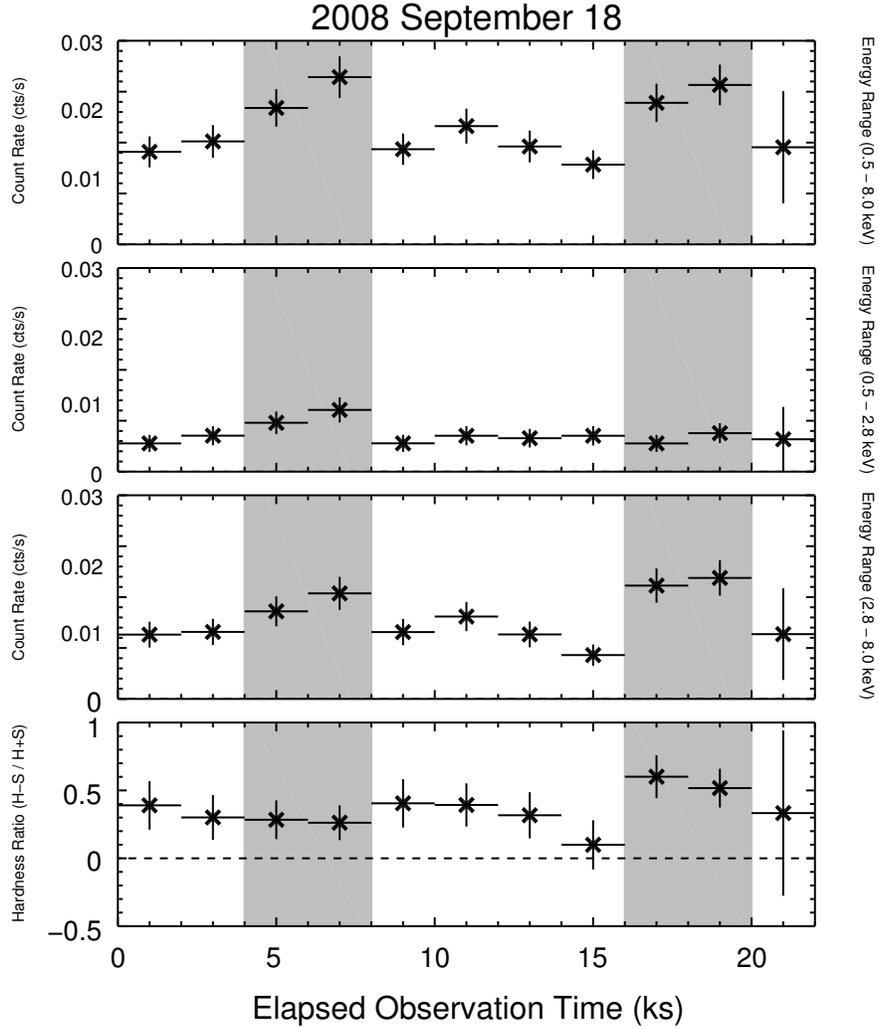}
\caption{X-ray light curves and hardness ratio of V1647 Ori during observation 9915.  The top panel light curve covers the 0.5--8.0 keV energy range, the second panel down spans the lower energy range of 0.5--2.8 keV,  and the third panel down covers the higher 2.8--8.0 keV energies.  The bottom panel displays the hardness ratio for each of the bins.  Counts were binned into 2 ks bins.  Uncertainties in mean hardness ratios and count rates are 1$\sigma$.  Apparent ``X-ray bright" time intervals are indicated by the shaded regions in the figure (4--8 ks and 16--20 ks) and in Figures \ref{10763hardness}--\ref{9917hardness}. \label{9915hardness}}
\end{figure}

\begin{figure}
\plotone{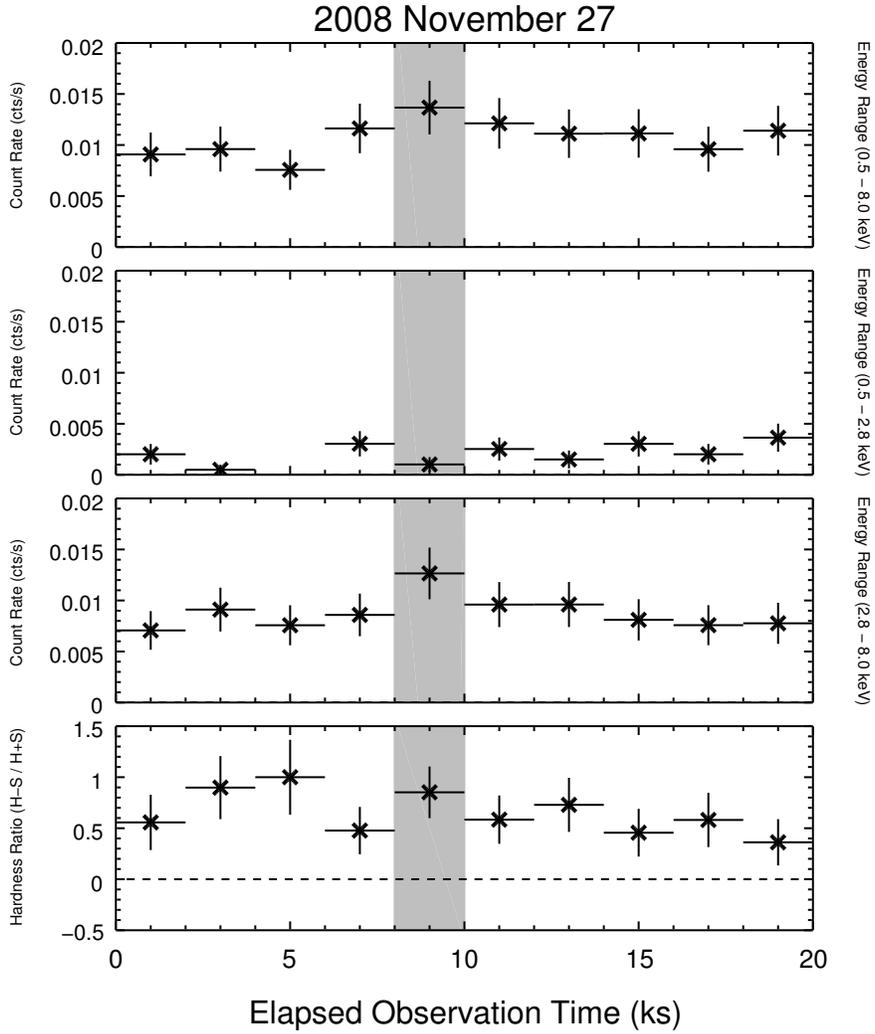}
\caption{X-ray light curves and hardness ratio of V1647 Ori during observation 10763.  See Figure~\ref{9915hardness} for description of panels.  Counts were binned into 2 ks bins. Uncertainties in mean hardness ratios and count rates are 1$\sigma$.  V1647 Ori had a slightly elevated X-ray flux, seen best in the hard X-ray band, from 8--10 ks.
\label{10763hardness}}
\end{figure}

\begin{figure}
\plotone{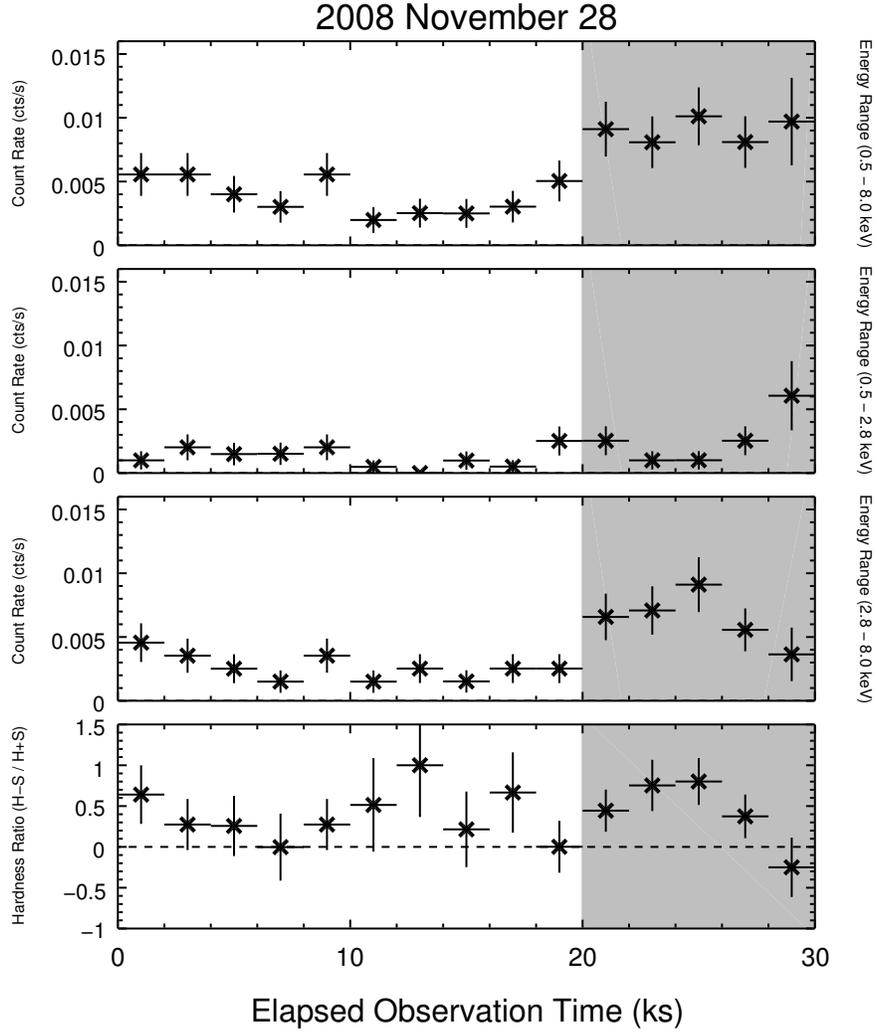}
\caption{X-ray light curves and hardness ratio of V1647 Ori during observation 8585.  Counts were binned into 2 ks bins.  Uncertainties in mean hardness ratios and count rates are 1$\sigma$.  The hard and broad-band X-ray light curves show that the hard X-ray flux from V1647 Ori increased sharply at 20 ks into the observation and remained elevated for $\sim$10 ks.  During this period of elevated X-ray flux, the soft-band flux remained more or less constant.  
\label{8585hardness}}
\end{figure}

\begin{figure}
\plotone{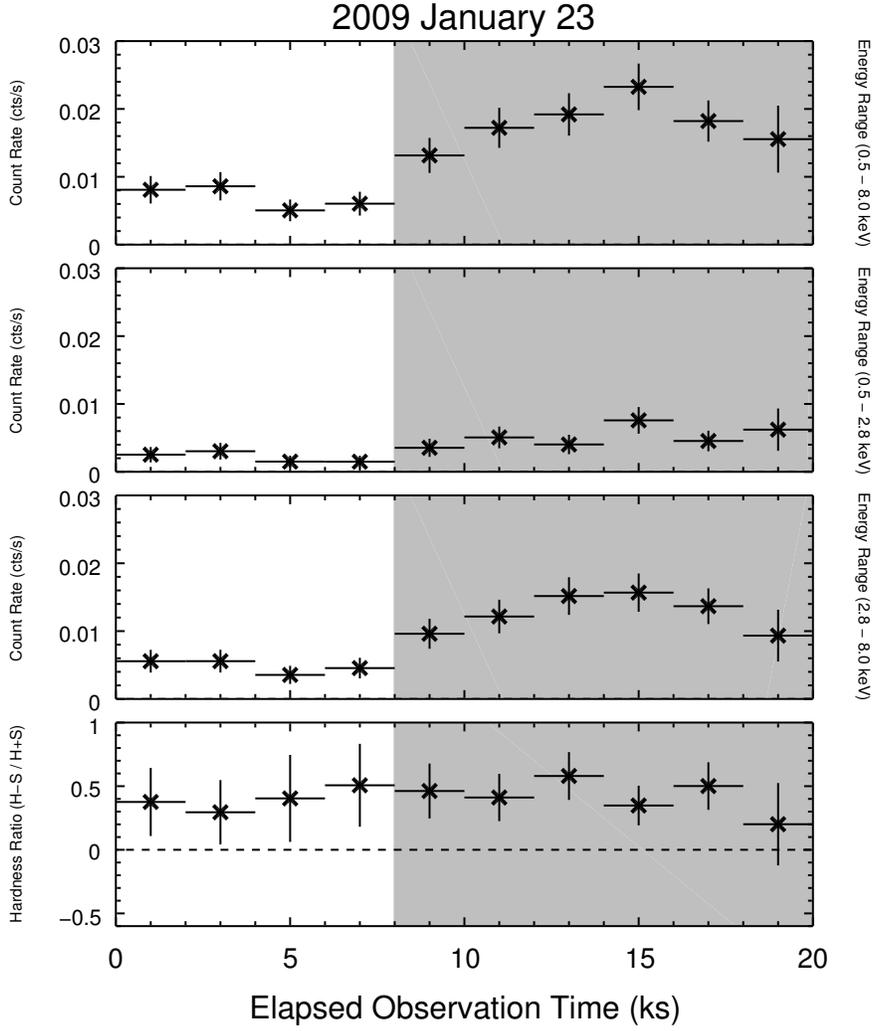}
\caption{X-ray light curves and hardness ratio of V1647 Ori during observation 9916.  Counts were binned into 2 ks bins.  Uncertainties in mean hardness ratios and count rates are 1$\sigma$.  The X-ray light curve shows that V1647 Ori began increasing in hard-band and, possibly, soft-band flux about 8 ks into the observation, and the flux remained elevated for $\sim$12 ks.  \label{9916hardness}}
\end{figure}

\begin{figure}
\plotone{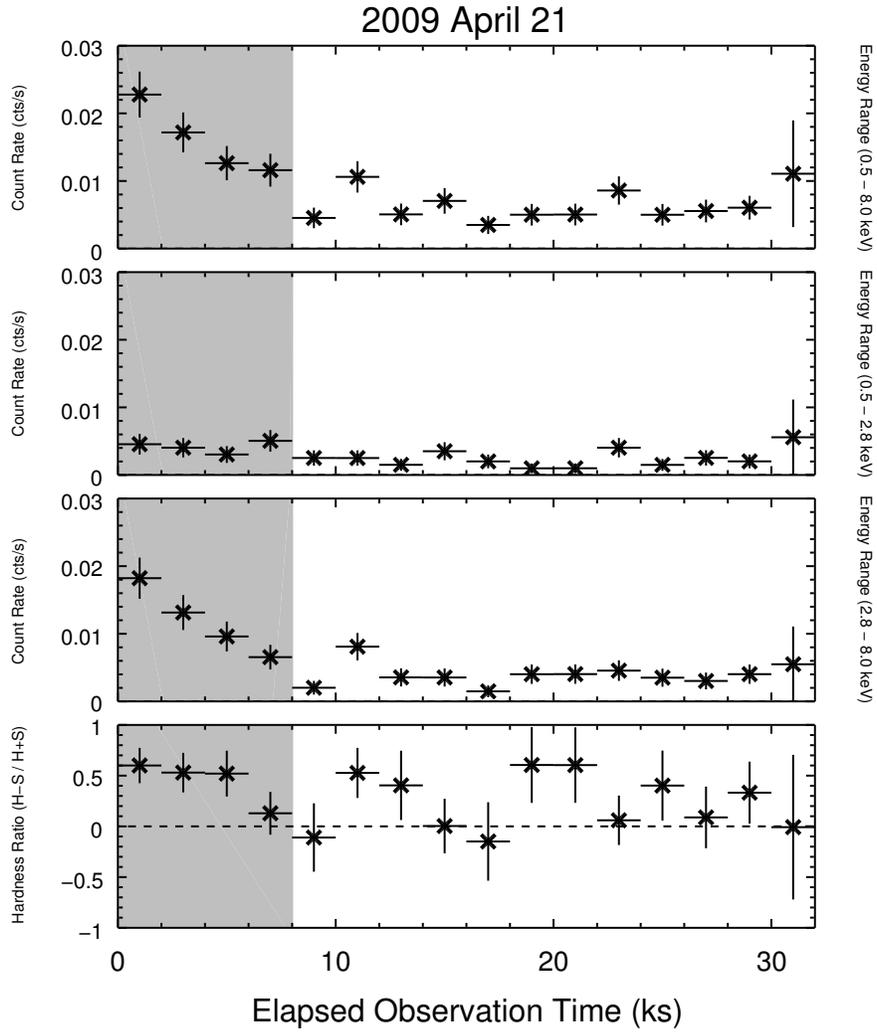}
\caption{X-ray light curves and hardness ratio of V1647 Ori during observation 9917.  Counts were binned into 2 ks bins.  Uncertainties in mean hardness ratios and count rates are 1$\sigma$.  V1647 Ori appeared to be in an elevated hard X-ray flux state at the onset of the observation, with the X-ray count rate remaining elevated for $\sim$8 ks.  \label{9917hardness}}
\end{figure}

X-ray light curves and hardness ratios as a function of time were extracted for each observation to examine the properties of any possible short-term X-ray variability.  In particular, these time series data permit us to investigate whether the trends in X-ray flux and hardness ratio seen in 2002--2006 were repeated during the recent outburst that began around 2008 August.  The extracted X-ray light curves and hardness ratio time series for six of the nine observations from 2002--2006 are presented in Figure~\ref{lcprev}.  X-ray light curves and hardness ratio curves for the five 2008--2009 Chandra observations are presented in Figures~\ref{9915hardness}--\ref{9917hardness}, which show (from top to bottom) the overall X-ray light curve, the soft (0.5--2.8 keV) light curve, the hard (2.8--8.0 keV) light curve, and the hardness ratio time series.  Energy ranges for the soft and hard X-ray bands follow those defined in \citet{gro05}, and plotted uncertainties in mean hardness ratios and count rates are 1$\sigma$.

During all five of the most recent (2008--2009) observations, V1647 Ori experienced very pronounced, short-duration ($\sim$2--10 ks) X-ray flux variations in the form of large rises and subsequent declines in count rate.  These fluctuations are evident at 4 ks and 16 ks in Figure~\ref{9915hardness}, at 8 ks in Figure~\ref{10763hardness}, at 20 ks in Figure~\ref{8585hardness}, at $\sim$8 ks in Figure~\ref{9916hardness}, and at the beginning of the observation reported in Figure~\ref{9917hardness}.  The mean count rate of V1647 Ori varied by about a factor of $\sim$2 for the first two epochs, while in the last three epochs, the mean count rate varied by a factor of $\sim$6.  In all five epochs, the bulk of the X-ray variability is contributed by the harder portions of the spectra.  

The X-ray count rate for V1647 Ori was intrinsically lower during most of the 2002-2006 observations (the count rate was $>$5 ks$^{-1}$ only on 2004 March 07) than in 2008--2009 (the count rate was $>$5 ks$^{-1}$ during all five observations).  Also, in some cases like ObsIDs 5307 and 5308 (both from 2004), the observations were on the order of only a few kiloseconds in duration.  The low count rates in most of the 2002--2006 data and the short exposure times in two of these observations make comparisons with the more recent datasets, in terms of the presence or absence of 2--10 ks timescale variations, fairly difficult.  However, observations 5382 and 5383, obtained in 2005 April and August, respectively -- more than a year after the initial optical outburst of V1647 Ori in 2003 November -- yielded enough counts that we can compare these two light curves with the more recent observations with some confidence.  As seen in these two light curves from 2005 (Fig.~\ref{lcprev}, lower-left panel, beginning at $\sim$10 ks, and lower-middle panel, beginning at $\sim$5 ks), the X-ray count rate again varied by a factor of $\sim$4 on a timescale of 2--10 ks, a level and duration of short-term variability very similar to that seen in the more recent (2008--2009) observations.  We conclude that kilosecond-timescale variations are seen in all observations of V1647 Ori in which the signal is strong enough and the observing window long enough for us to search for such events, and that the amplitude duty cycle and duration of these few-kilosecond X-ray variations were stable over the five-year period from 2005 to 2009.  Such variability appears to be part of the normal X-ray pattern for this object, at least when the count rate is elevated.  One therefore needs to interpret the overall flux levels observed during the 2004 March 07 and 2004 March 22 observations, which, respectively, are only 5.5 ks and 4.9 ks in duration, with caution.

The highest measured X-ray mean count rate following the latest (2008) optical outburst of V1647 Ori is higher (23.6 ks$^{-1}$; one month post-outburst) than the highest level (11.7 ks$^{-1}$; four months post-outburst) observed during its 2003--2005 outburst (Table~\ref{tab-1}).  In addition, with the exception of the 2004 March 07 observation, the 2008--2009 observations (a period spanning seven months) show V1647 Ori to be consistently brighter than it was during the 2003--2005 outburst.  We find that in 2008--2009 the X-ray spectrum stays relatively hard, as is shown by the median photon energies, which in 2008--2009 are consistently about twice the energy (3.4 to 3.9 keV) that was observed before the first eruption (2.0 keV in 2002) and after V1647 Ori had returned to its quiescent state in late 2005 (1.7 keV).  Also, while the mean hardness ratios were negative both before (2002 November) and after (2005 December) the first eruption, the mean hardness ratios in 2008--2009 stay consistently around $\sim$0.3 or greater. 

These results also reveal that V1647 Ori exhibited a very interesting sequence of changes of state, beginning in 2004 March.  V1647 Ori had switched from its elevated state on 2004 March 07 (11.7 counts  ks$^{-1}$, median photon energy of 3.6 keV, mean hardness ratio of +0.46) to what appears to be a quiescent state on 2004 March 22 (2.5 cts ks$^{-1}$, median photon energy of 2.0 keV, $-$0.45 mean hardness ratio) but then was again in an elevated state one year later on 2005 April 11 (4.8 cts ks$^{-1}$, median photon energy of 3.5 keV, mean hardness ratio of +0.51).  Yet by 2005 August 27, V1647 Ori was again returning to quiescence (0.9 cts  ks$^{-1}$, 3.0 keV median photon energy, +0.01 mean hardness ratio) and by 2005 December was fully back to quiescence.    

At least three explanations are plausible for the observed changes we have described in the X-ray emission from V1647 Ori that occurred between 2004 March 07 and 2004 March 22 and between 2004 March 22 and 2005 April 11.  One:  the outbursts generate both hard and soft X-rays; most of the time (2004 March 07 and 2005 April 11), the hard X-ray plasma is dominant in the X-ray spectrum as observed by CXO, but at certain times (2004 March 22) the soft X-ray plasma component is seen more clearly.  Two:  the first major outburst (2003 November through early 2004 March) had ended, at least in the X-ray regime, after only a few months.  By 2005 April, the source was again in an elevated state and by 2005 August, four months later, was again returning to quiescence.  Three:  the 2004 March 22 observation caught V1647 Ori during a short duration (2--10 ks) quiescent moment during the extended (months long) outburst phase.  However, the XMM-Newton observation obtained on 2004 April 4 showed a low X-ray flux level at the beginning of the observation that is consistent with the quiescent level observed by CXO that increased in the second part of the observation \citep{gro05}.  Moreover, the supporting evidence from observations made at optical and near-infrared wavelengths for the first outburst ending in 2004 mid-March and for a second outburst beginning later that year (explanation ``Two") is weak, at best.  Of the remaining two possibilities, we will argue in \S4 that explanation ``One" is more likely.

\begin{figure}
\centering
\epsscale{1.0}
\plottwo{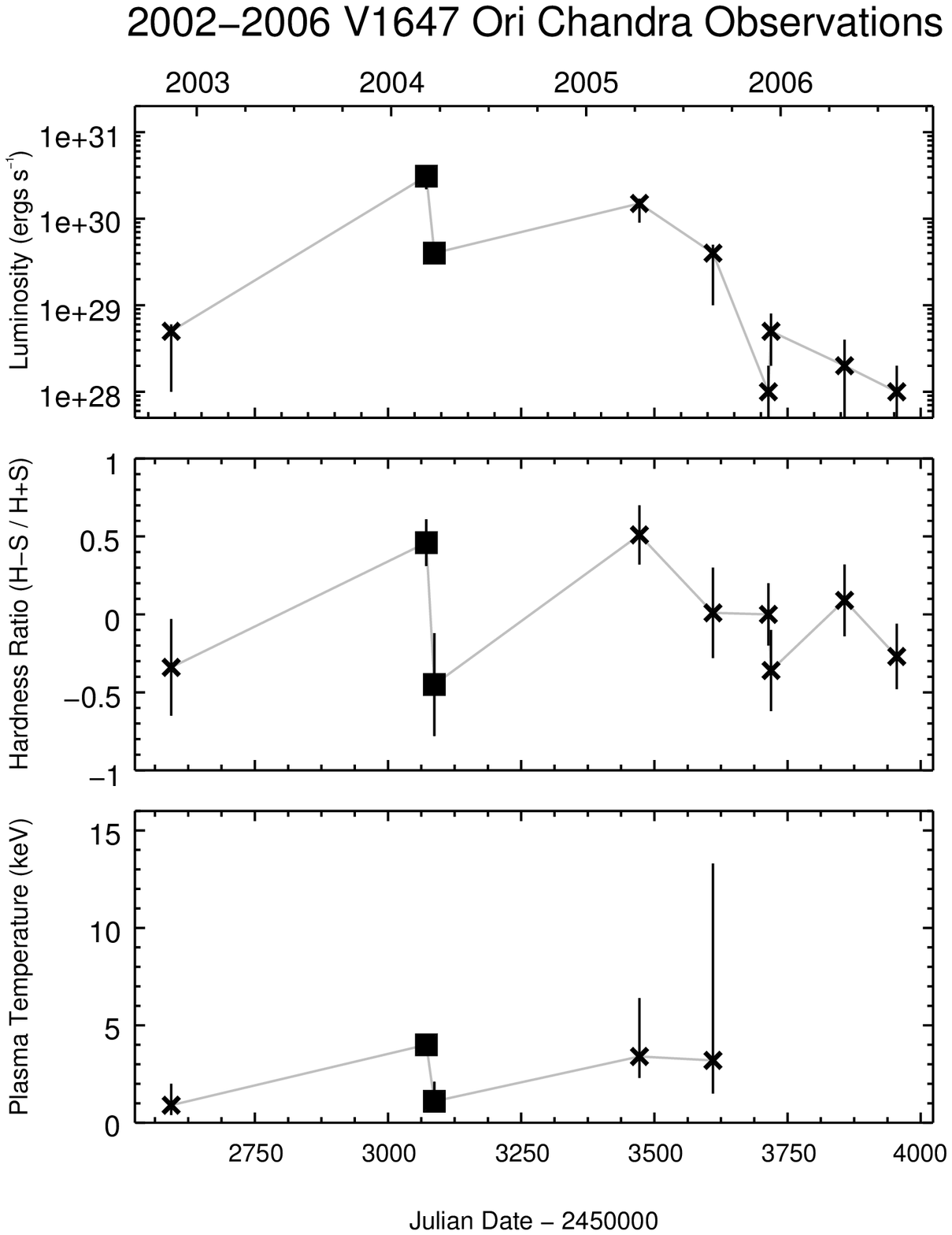}{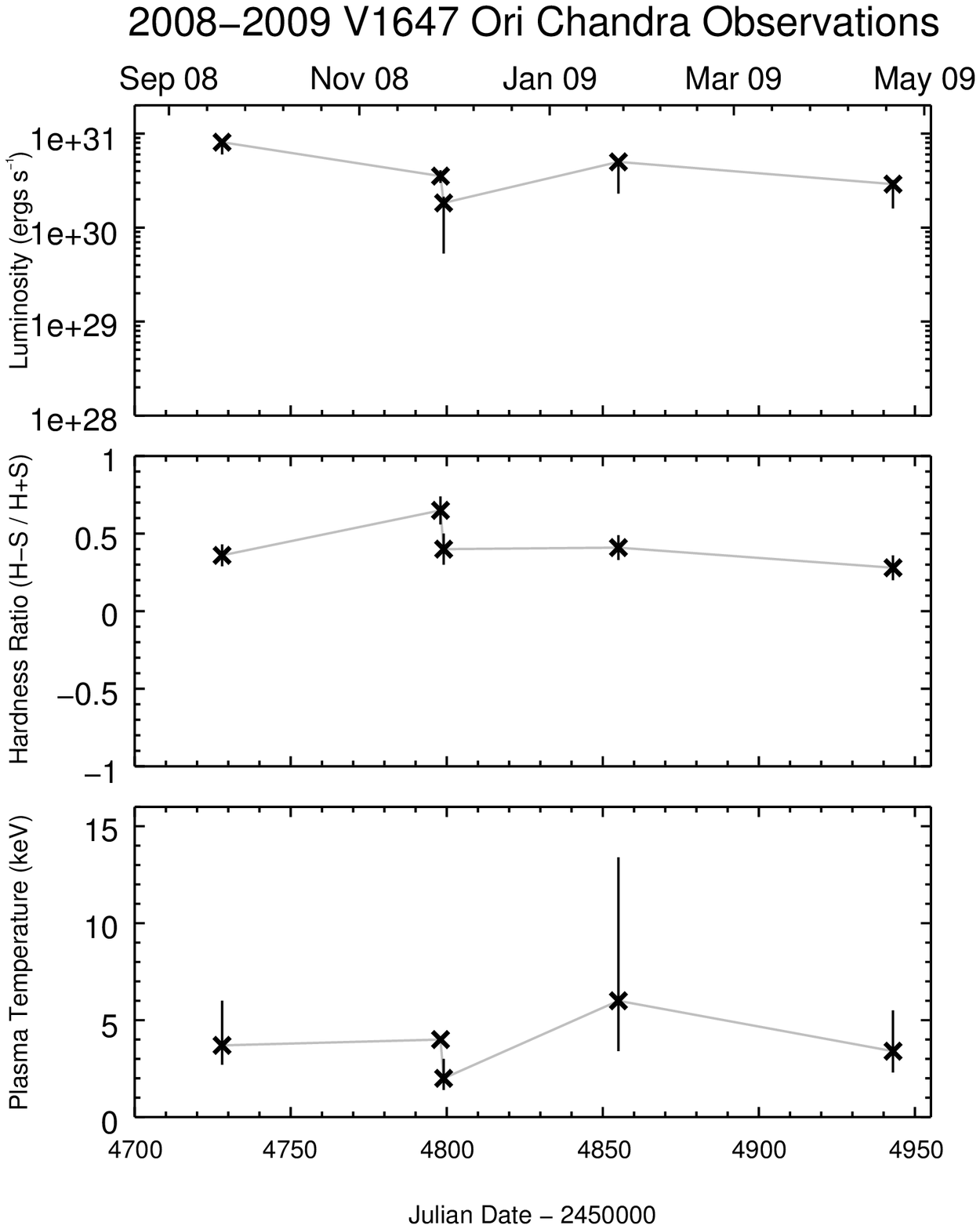}
\caption{Time series of observed X-ray luminosity (top panels), mean hardness ratio (middle panels), and plasma temperature (bottom panels) for V1647 Ori.  Crosses represent data obtained with ACIS front-illuminated CCDs, and squares represent data obtained with the ACIS back-illuminated S3 CCD. Plotted uncertainties for the hardness ratios and luminosities represent the 68\% confidence interval (1$\sigma$) and plotted uncertainties for the plasma temperatures represent the 90\% confidence interval (1.6$\sigma$).  The modeled plasma temperatures for the 2005 December through 2006 May observations are not shown because this parameter was not well constrained by spectral fitting.
\label{V1647OriLumHRTemp}}
\end{figure}

\begin{figure}

\plotone{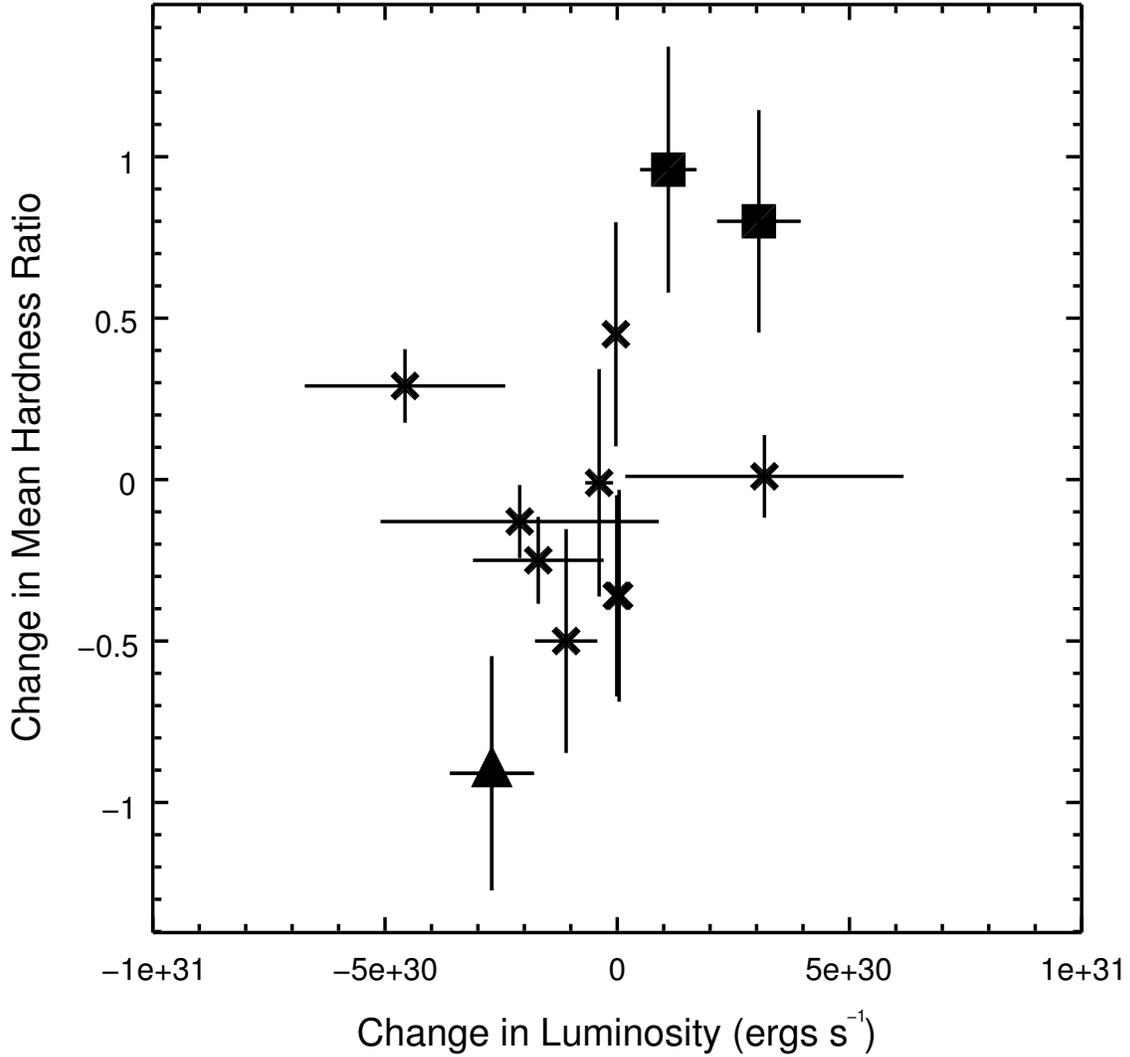}
\caption{Two-parameter plot for all CXO observations of V1647 Ori showing the correlation between the changes in mean hardness ratios of individual observations and the changes in observed X-ray luminosities.  Crosses represent values corresponding to data obtained with ACIS front-illuminated CCDs, squares represent values involving data obtained with ACIS front-illuminated CCDs and the back-illuminated S3 CCD, and the triangle represents a value that used back-illuminated S3 CCD data only.  The correlation coefficient for the changes in mean hardness ratio and changes in observed X-ray luminosity is 0.44.  \label{longterm_corr}}
\end{figure}

A comparison of the count rates (Figures \ref{9915hardness}--\ref{9917hardness}, top panels) and hardness ratios (Figures \ref{9915hardness}--\ref{9917hardness}, bottom panels) during the $\sim$2--10 ks variability periods suggests that the X-ray spectrum of V1647 Ori may harden as it brightens and soften as the source fades during these few-kilosecond variations. This appearance of a correlation between the changing count rate and the changing hardness ratio on a few kilosecond timescale during short-term variability, however, is not consistent.  In fact, we find a weak anti-correlation between these properties (correlation coefficient = $-$0.18) on these short timescales; however, we do find a strong, positive correlation over much longer periods of time between the X-ray luminosity (calculated in our spectral modeling; see Section \ref{modeling}) and hardness ratio, as shown in Figure~\ref{V1647OriLumHRTemp}. In this figure, the top panels show the seven-year X-ray luminosity curve of V1647 Ori while the middle panels show the corresponding hardness ratios with a single, average measurement plotted for each observation epoch. A more direct demonstration of this correlation is shown in Figure  \ref{longterm_corr} where we plot the change in mean hardness ratio versus the change in X-ray luminosity between observations of V1647 Ori. We derive a correlation coefficient of 0.44 for these two parameters.

As seen in the top and middle left-hand panels of Figure~\ref{V1647OriLumHRTemp}, from the first Chandra observation in 2002 to the first observation of V1647 Ori in 2004, the luminosity increased by $\sim$2 orders of magnitude while the mean hardness ratio increased from a relatively soft value of $-$0.34 to a hard value of 0.46.  From the second to the third observation, the X-ray luminosity dropped by an order of magnitude while the mean hardness ratio softened to a much lower value of $-$0.45, much lower than that seen in the 2002 observation of V1647 Ori when it was seen to be in quiescence. When the fourth Chandra observation occurred in 2005, the luminosity had risen back to outburst levels again while the mean hardness ratio increased again to +0.51, the highest observed to that date.  In subsequent Chandra observations through early 2006, the X-ray luminosity of V1647 Ori diminished while the hardness ratio dropped.

In the top and middle righthand panels of Figure~\ref{V1647OriLumHRTemp}, we present the X-ray luminosity and mean hardness ratio versus time for most recent Chandra observations of V1647 Ori.  The same correlation between changes in X-ray luminosity and changes in the mean hardness ratio seen in the 2002 through 2006 observations is evident in all observations from 2008 through 2009.  From 2008 September 18 to 2008 November 27, there appears to be an anti-correlation with the luminosity decreasing slightly between these observations while the hardness ratio increases.  We note, though, that in both of these epochs, the luminosity is high and the spectrum is quite hard.  Since the short-period light curves of each observation show that there is typically X-ray variability on timescales of kiloseconds, with X-ray flux sometimes increasing by several factors, this apparent anti-correlation might be attributed to short-duration variability.

These results suggest that we can identify two states for V1647 Ori, as seen in time-averaged data on a 5--30 ks timescale:
\begin{itemize}
\item In its \emph{quiescent} state, V1647 Ori has a low count rate ($<$1 ct ks$^{-1}$), low median photon energy ($\lesssim$2.5 keV), and negative (soft X-ray dominated) hardness ratio.  
\item In its \emph{elevated} state, the count rate is at least 25--50 times greater, the median photon energy doubles, and the hardness ratio becomes strongly positive (hard X-ray dominated).  
\end{itemize}
The X-ray evidence shows that when V1647 Ori enters a major optical/near-infrared outburst, the X-ray profile switches from quiescent to elevated, and when the outburst phase ends, the X-ray profile switches from elevated to quiescent. 

\subsection{Spectral Modeling}\label{modeling}

\subsubsection{Procedure \& Results}\label{singletemp} 

Modeling of the spectra employed XSPEC v12.4.  For those spectra with single-count binning, XSPEC was set to use the Cash-statistic instead of the $\chi$$^{2}$ statistic to assess the goodness of fit.  All models at first assumed a thin, single-temperature plasma (APEC model) subject to absorption by an intervening column of hydrogen (WABS component).  From the 2004 April XMM-Newton observations, single-component thermal plasma models of the V1647 X-ray spectrum yield a hydrogen column density, chemical abundance, and plasma temperature of N$_{H}$ = 4.1$\times$10$^{22}$ cm$^{-2}$, Z = 0.8 solar, and kT$_{X}$ = 4.2 keV, respectively \citep{gro05}, and recent Suzaku observations yield similar results \citep{ham10}.  Therefore, initially, N$_{H}$, Z, and kT$_{X}$ were set approximately to the aforementioned values (N$_{H}$ = 4.0 $\times$ 10$^{22}$ cm$^{-2}$, Z = 0.8 solar, and kT$_{X}$ = 4.0 keV).  As our initial test indicated that Z is poorly constrained by the Chandra data, only N$_{H}$ and kT$_{X}$ were left free to vary during the fitting procedure.  For four of the five 2008--2009 observations, the models converged to best-fit solutions with column density and plasma temperature within a 90\% confidence interval of their initial values; however, the 2008 November 27 spectrum model was unable to converge to a physically-meaningful fit, so the plasma temperature was fixed at kT$_{X}$ = 4.0 keV (Table \ref{tab-3}).  

 A similar fitting method was used for the 2002--2006 observations (Table \ref{tab-2}).  Initially, the first five of these spectra spectra were modeled with freely varying hydrogen column densities and plasma temperatures, while the chemical abundance was fixed at 0.8 solar.  These spectral models, however, were unable to constrain plasma temperatures or X-ray fluxes and luminosities, so the hydrogen column density was then fixed at N$_{H}$ = 4.1 $\times$ 10$^{22}$ cm$^{-2}$, and the models were refit.  The remaining four spectra, obtained when V1647 Ori was reverting to an optical/X-ray quiescent state, were modeled with the hydrogen column density and plasma temperature fixed at N$_{H}$ = 4.1 $\times$ 10$^{22}$ cm$^{-2}$ and  kT$_{X}$ = 0.86 keV, respectively.  Given the small number of counts in these four spectra, the associated error ranges for these X-ray fluxes and luminosities were obtained by multiplying the error ranges for the mean count rates of these observations with an appropriate energy conversion factor (ECF), where the ECF for each observation epoch was obtained by dividing the derived X-ray flux by the mean count rate.
 
For the CXO observations of V1647 Ori with sufficient total counts (ObsIDs 9915 and 9917), we performed fits of a two-component thermal plasma model with parameters for hydrogen column density set to 4.0 $\times$ 10$^{22}$ cm$^{-2}$, plasma temperatures set to kT$_{X}$ = 0.5 keV and 2.0 keV, and chemical abundance fixed at 0.8 solar.  Visual inspection of the spectral models shows that there is a negligible difference between the best-fit single- and two-component models, F-test results suggest that there is no statistical improvement in the model fits with the addition of a second plasma component; i.e., the best-fit parameters for the latter model converge on values such that the contribution of the lower-temperature component is negligible.  We conclude, therefore, that all of the 2008--2009 CXO data are best fit with a single-component model.

Best-fit models for each of the five recent Chandra observations are shown in Figure~\ref{bestfive}.  The overall trend of the spectral models is to converge to fits with parameters similar to those of the best-fit single-component model reported by \citet{gro05}.  Intervening hydrogen column densities do vary from model to model but remain in the N$_{H}$ $\sim$ 4--6 $\times$ 10$^{22}$ cm$^{-2}$ range, and plasma temperatures are kT$_{X}$ $\sim$ 2--6 keV, slightly higher than but still within the range of uncertainty found by \citet{kas06}. 

Determining a robust model for the last three recent epochs was more challenging than for the first two epochs, especially for ObsID 9916 (2009 January 23). A major concern for us was to assess whether the flux and hardness variability seen in the light curves for these observations (Figs. \ref{8585hardness}--\ref{9917hardness}) contributes to the difficulty of fitting a unique model to the spectral data. To assess the effects of variability on the robustness of our spectral analysis, we divided each of these the spectra into two phases, one in which the X-ray flux was relatively lower and one in which the flux was elevated. Each of these spectra was then modeled independently. Though the hardness ratios suggest that the spectra did harden marginally as V1647 Ori went from X-ray-dim to X-ray-bright during these observations, the spectral modeling yields no measurable changes in hydrogen column density or plasma temperature.  Thus, it appears that the changes in the light curves are likely due only to changes in the plasma emission measure.

\subsubsection{Iron Line Emission Analysis}\label{ironline} 
Our initial modeling of the V1647 spectra from 2008 and 2009 suggests the presence of line emission from near-neutral iron at 6.4 keV as well as from the (unresolved) helium-like iron K$\alpha$-line triplet at 6.64, 6.67, and 6.70  keV.  The neutral iron line emission at 6.4 keV is often seen in accretion-powered sources and is usually attributed to the fluorescence of cold gas in the presence of a nearby X-ray continuum emission  \citep{tsu05}.  Thus, the presence of this line in some of the V1647 Ori spectra could be attributed to fluorescence of (neutral) circumstellar disk material by accretion-generated X-rays.  It would not be surprising to detect this emission in the spectra of V1647 Ori given that the environment of this YSO does appear to contain the necessary ingredients for the formation of 6.4 keV emission, namely a strong, relatively hard X-ray source illuminating cold circumstellar material.  Observations of other YSOs, including roughly a half-dozen sources in the COUP survey, with these environmental components have also shown this feature.

\begin{figure}
\centering
\includegraphics[scale=0.65]{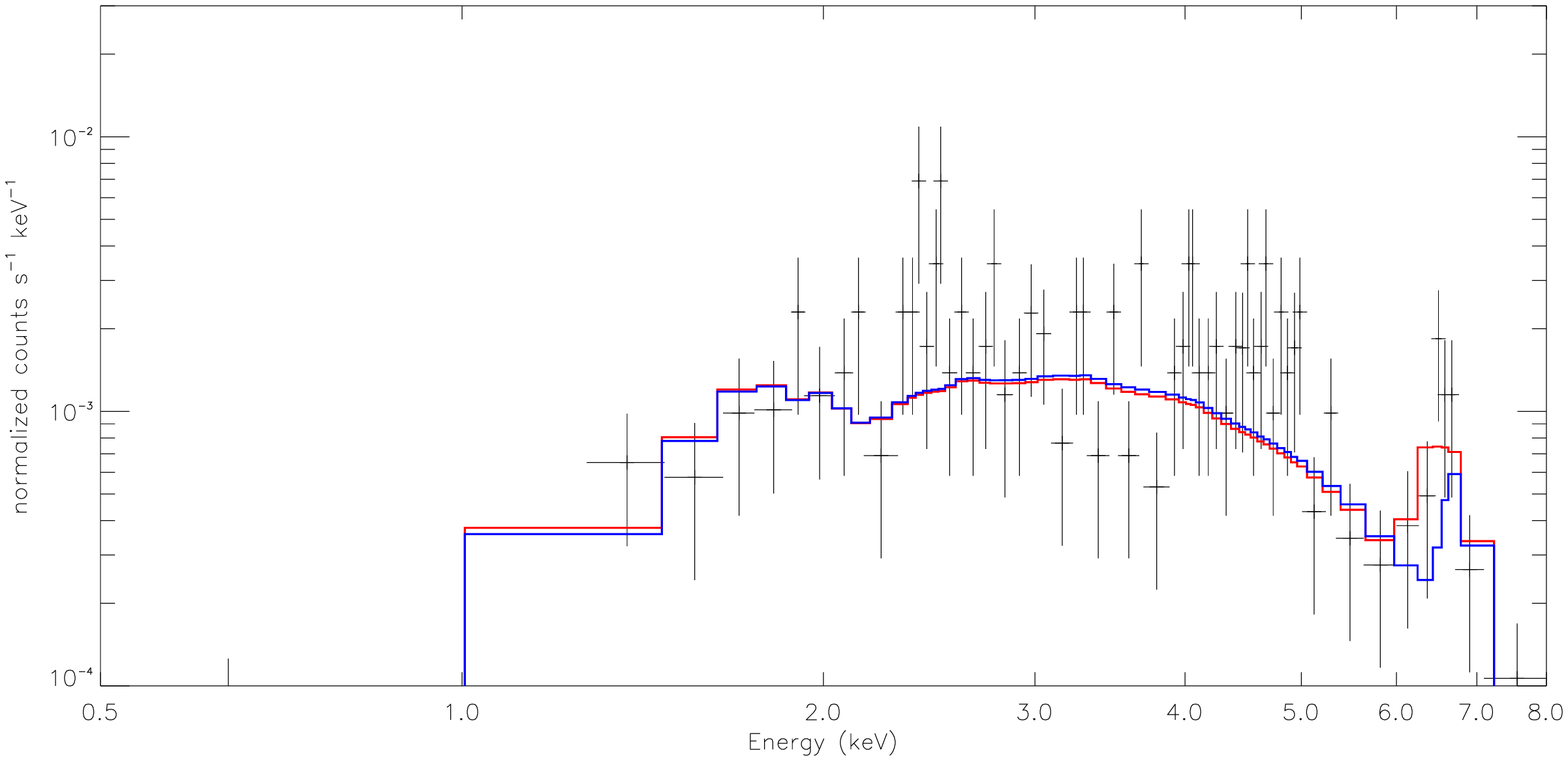}\\  
\includegraphics[scale=0.65]{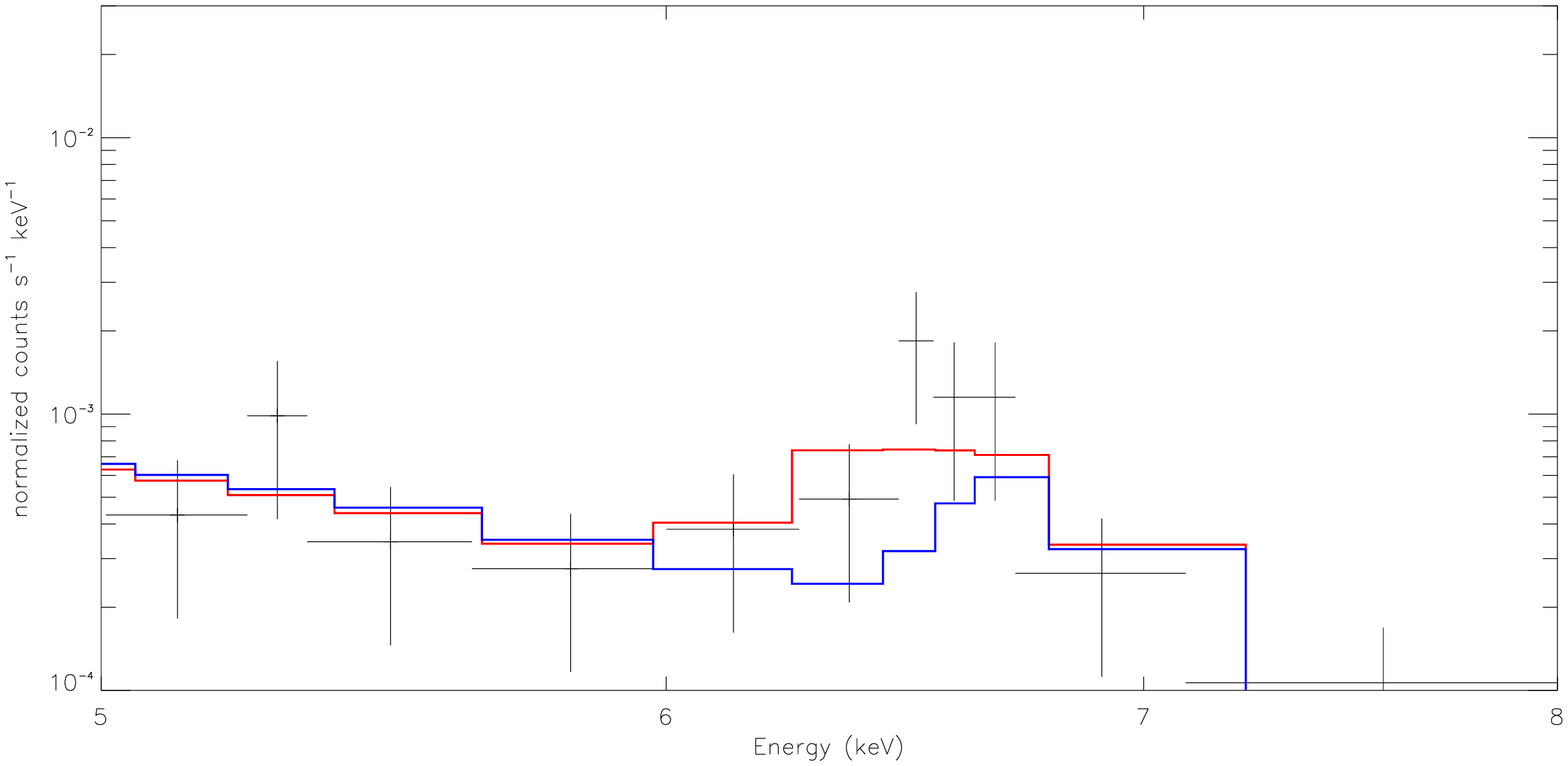}\\ 
\caption{Spectrum of the 2009 April 21 observation (black) overlaid with the single-component plasma model (blue) and the model with the additional Gaussian component (red) centered at 6.4 keV (equivalent width of 510 eV) added to account for neutral iron emission.  The entire modeled wavelength range (0.5--8.0 keV) is shown in the top panel, and the bottom panel displays the 5.0--8.0 keV energy range to more easily show the iron emission and the Gaussian component.  Spectral data points employ a three-count-minimum bin size.\label{irongaussian}}
\end{figure}

Following \citet{gro05}, we added a Gaussian component centered at 6.4 keV to the spectral models in order to account for the neutral line component.   Models of two observations (2008 September 18 and 2009 April 21) appear to be well fit with the addition of a 6.4 keV line with equivalent widths of $\sim$200 and $\sim$500 eV, respectively, while the 2008 November 27-28 and 2009 January 23 are fit well without the additional neutral iron line emission component (Fig. \ref{irongaussian}).  On the other hand, it is possible that the 6.4 keV emission is present during all of the 2008--2009 observations from CXO but that its spectral signature is muffled by noise in the 2008 November and 2009 January observations.  While it appears that we have detected 6.4 keV iron emission in the 2008 September 18 and 2009 April 21 observations, we cannot conclude definitively that these ``detections" are real.  We note that the best-fit 6.4 keV equivalent widths are poorly constrained with error ranges extending from zero to roughly twice the equivalent width values.  Given this range of uncertainty, we cannot exclude the possibility that there is no 6.4 keV iron line.  We feel confident, however, that these detections are real given that visual and quantitative comparisons of our findings with those detections found by \citet{tsu05} are very similar.  

We have compared the intensities/appearances of this line in the various observations of V1647 Ori by CXO, XMM-Newton, and Suzaku.  The equivalent widths of this emission feature are very similar to those found in the spectra of the 2004 April 4 XMM observation (109 eV) and in the 2008 October 8 Suzaku observation ($\sim$600 eV).  We are unable to clearly determine whether there is a correlation between the strength and appearance of the 6.4 keV line and any of the associated plasma characteristics.      

\section{Discussion}\label{discussion}

\begin{figure}
\epsscale{1.0}
\plotone{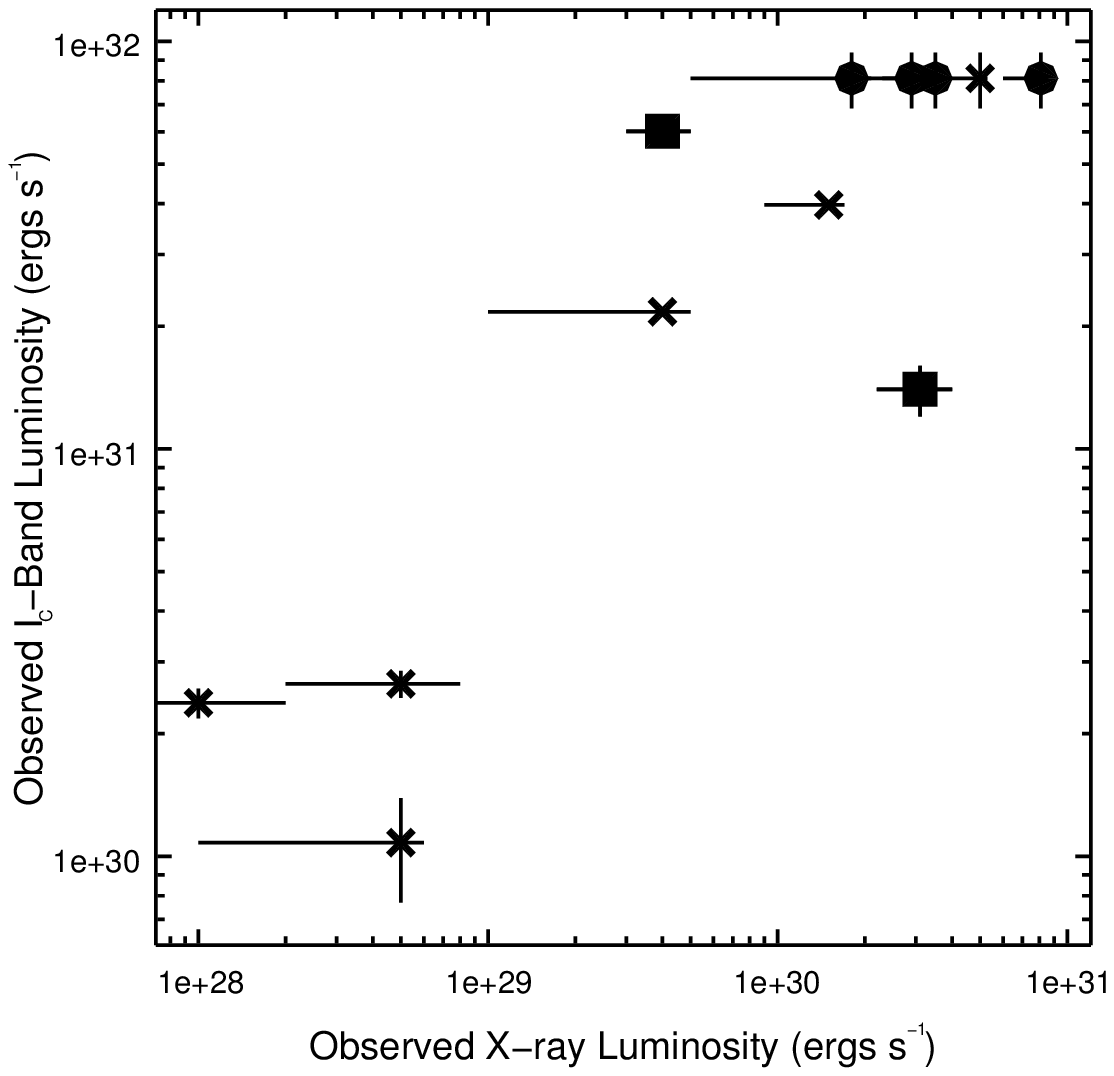}
\caption{Two-parameter plot for all CXO observations of V1647 Ori showing the correlation between the observed I$_{c}$-band near-infrared and X-ray luminosities.  Crosses represent data obtained with ACIS front-illuminated CCDs, and squares represent data obtained with the ACIS back-illuminated S3 CCD.  Errors for I$_{c}$-band luminosities that were interpolated at X-ray observation dates were calculated by averaging the errors of the I$_{c}$-band luminosities of the five observations made nearest to the interpolation date.  Circles represent four observations (ObsIDs 9915, 10763, 8585, and 9917) in which the interpolated I$_{c}$-band luminosity of ObsID 9916 was assumed to be the correlated I$_{c}$-band luminosity. The correlation coefficient for the  I$_{c}$-band and X-ray luminosities without the circle values is 0.65, while the correlation coefficient using all values is 0.67.  \label{longtermhvsx}}
\end{figure}

Figure \ref{overall} suggests that the overall X-ray flux of V1647 Ori is strongly correlated with optical/near-infrared flux.  This correlation is revealed more clearly in Figure \ref{longtermhvsx}, in which we plot the X-ray luminosity versus the I$_{C}$-band luminosity.  We interpolated I$_{C}$-band luminosities for the 2002--2005 X-ray observation dates.  We did not extrapolate I$_{C}$-band luminosities for the two 2006 X-ray observation dates due to the highly uncertain flux behavior of V1647 Ori.  Of the 2008-2009 X-ray observations, we could only interpolate an I$_{C}$-band luminosity for ObsID 9916.  With these eight interpolated luminosities, we derived a correlation coefficient between X-ray luminosity and I$_{C}$-band luminosity of 0.65.

Dramatic increases in optical/near-infrared flux for YSOs, such as FU Ori, have long been thought to be associated with enhanced accretion \citep{har96}.  In such an environment, material is channeled through magnetic funnels from the co-rotation radius of the circumstellar disk down to the photosphere \citep{shu94}.  Hence, when the X-ray flux from a pre-main-sequence star or protostar is elevated and the rapid rise in X-ray emission is directly correlated with large-scale optical outbursts, the correlation itself strongly suggests that accretion is the mechanism responsible for generating the increase in X-ray output \citep{kas06}.  

One way in which accretion-generated X-ray emission could be identified observationally would be through the relatively soft X-rays emitted by the plasma when it plunges onto the stellar surface at free-fall velocities and is shock heated to temperatures of a few million Kelvin (kT$_{X}$ $\sim$0.3 keV) \citep{kas02,ste04,raa09,brick10}.  We can speculate that a second such signature would be much hotter (kT$_{X}$ $\sim$few to tens of keV) and harder X-rays generated in magnetic reconnection events \citep{shi02,brick10} within the accretion streams.  Whether the soft or hard X-ray generating plasma is observable likely depends on the observing geometry.  If we have an unobscured line of sight to the footprint of the accretion column, our observations should be sensitive to the cooler plasma; if the accretion column obscures our view of the accretion-column footprint, our observations should be sensitive only to the hotter plasma in the accretion stream; and if the obscuration of the accretion footprint is partial, we might detect both plasmas.  In between periods of dramatically enhanced accretion, previous studies such as COUP \citep{pre05,sta06} and the XMM-Newton Extended Survey of the Taurus molecular cloud (XEST) \citep{aud07}, suggest that the X-ray signature of young stars should be that of normal coronal emission.  Such emission would be similar to but much fainter than the hot, hard plasma seen from reconnection events in the accretion funnel. 

During quiescence, the X-ray flux of V1647 Ori has a low count rate, low median photon energy, and negative hardness ratio; in contrast, when the X-ray flux is elevated, the count rate increases by a factor of 25 or greater, the median photon energy doubles, and the hardness ratio becomes strongly positive.  Most of the V1647 Ori light curves reveal that this YSO also experiences what appear to be short-term (few kilosecond) variability in its X-ray flux.  

\subsection{Observations of Hotter and Cooler X-ray-Generating Plasmas}

Since most of the CXO observations show this short-duration variability when the X-ray count rate is high, the short-term variations are likely part of the normal behavior for V1647 Ori.  For all 2008--2009 CXO observations, the X-ray variability is almost entirely seen in hard ($>$2.8 keV) X-rays.  Our model fits of the V1647 Ori spectra indicate that the X-ray signature of the plasma we observe from this source during optical eruptions is predominantly a bright, hot, hard, single-temperature plasma heated to temperatures of 2--6 keV, which is also consistent with X-rays generated from reconnection events in an accretion funnel \citep{iso03,brick10}.  

We must keep in mind, however, that the intervening hydrogen column density will play a significant role in our ability to detect X-ray flux from accretion hotspots at temperatures of a few million Kelvin, especially when dealing with column densities as high as those modeled in V1647 Ori observations.  An accretion-footprint (shocked) plasma at a temperature of kT$_{X}$ $\sim$ 0.3 keV subject to an intervening absorbing column density of a few times 10$^{22}$ cm$^{-2}$ would have $\sim$98\% of its X-ray flux extinguished.  Even given the smallest hydrogen column density found via spectral modeling of the 2004--2005 V1647 Ori observations when the hydrogen column density was allowed to vary freely ($\sim$1 $\times$ 10$^{22}$ cm$^{-2}$), such a soft component still has $\sim$96\% of its X-ray flux absorbed.  If the hydrogen column density decreased dramatically --- to N$_{H}$ of a few times 10$^{21}$ cm$^{-2}$ --- the accretion shock emission could dominate the observed flux.  This seems a plausible scenario to explain the softer plasma detected in late 2004 March, especially if the large intervening hydrogen column density inferred at other observing epochs was due mostly to the accretion streams. 

An alternative explanation is that we observed V1647 Ori when it was in the midst of a large accretion episode that pushed the star-disk boundary inward to the point where the accretion became non-magnetospheric \citep{har98}, effectively reducing the amount of hard X-ray flux produced by magnetic reconnection events in the accretion stream.  

During the 2008--2009 epoch, when the X-ray luminosity of V1647 Ori increased, the spectrum hardened and the emitting plasma increased in temperature (Fig.~\ref{V1647OriLumHRTemp}, right panel); also, when the overall X-ray luminosity decreased, the X-ray spectrum softened and the X-ray generating plasma cooled.  These correlations are also seen in the Chandra observations following the 2003 eruption (Fig.~\ref{V1647OriLumHRTemp}, left panel).  However, between 2008 September and 2008 November, the spectrum appeared to harden slightly as the X-ray luminosity decreased slightly.  If we are observing X-rays generated predominantly by the $\sim$1 keV plasma in the accretion footprint, then a decline in accretion should result in a decline in total X-ray luminosity, particularly in the lower energy flux, resulting in a hardening of the spectrum.  Therefore, the hardening of the spectrum and decrease in X-ray luminosity observed between 2008 September and November might be a time when CXO was able to see plasma closer to the protostellar surface.  In fact, this epoch showed a low temperature ($\sim$2 keV) plasma, consistent with what could be a mixture of hotter (4--6 keV) and cooler (1 keV) plasmas.  These data may offer evidence that the X-ray flux includes an emission component from the cooler plasma in the footprint of the accretion funnel located in the stellar photosphere.  Because the cooler plasma suffers greater extinction and because CXO is less sensitive to softer X-ray emission than XMM-Newton, CXO would likely only detect the cooler plasma when the viewing geometry is favorable.  

\subsubsection{Intervening Hydrogen Column Density}

Our modeling work for the X-ray observations obtained in 2004--2005 and 2008--2009, when the mean X-ray count rates were greater than 1 cts ks$^{-1}$ yield a best fit value for $N_{H}$ of about 4.1 $\times$ 10$^{22}$ cm$^{-2}$, consistent with the results derived by \citet{gro05} and a visual extinction of $A_{V}$ $\sim$ 20 \citep{vuo03}. We were unable to fit  $N_{H}$ in our modeling work for the low count-rate observations during the quiescent period in 2005--2006; however, \citet{asp08} obtained a best fit value for $A_{V}$ of 19 $\pm$ 2, based on their optical, near-infrared, and mid-infrared observations obtained in February 2007, which was also during the quiescent period (based on the optical and near-infrared photometry reported by \citet{asp08}). The observed value of $A_{V}$ for February 2007 lends strong support to our use of $N_{H}$ = 4.1 $\times$ 10$^{22}$ cm$^{-2}$ for our modeling of the quiescent epoch observations.  In addition, together these data suggest that $N_{H}$ and, by implication, $A_{V}$,  remained essentially unchanged as inferred from the X-ray observations, whether V1647 Ori was in the quiescent or elevated X-ray state.

On the other hand, as seen at longer wavelengths, the extinction toward V1647 Ori has changed.   \citet{abr04} derive $A_{V}$ = 13 from near-infrared data obtained in 1998 by 2MASS. During the outburst in 2004--2005, \citet{bri04} found $A_{V}$ = 8--10 on 2004 February 18, \citet{rei04} found $A_{J}$ = 1.26, $A_{H}$ = 0.81, and $A_{K}$ = 0.5 on 2004 Feburary 18, all of which are consistent with $A_{V}$ of $\sim$5 \citep{bec78}, \citet{vac04} obtained $A_{V}$ = 11 from measurements of the 3.1 $\mu$m water band on 2004 March 9, and \citet{ojh06}, who made optical and near-infrared observations from 2004 into very late 2005, reported $A_{V}$ of $\sim$5 during the 2004 outburst.  \citet{ojh06} then report that $A_{V}$ increased to $\sim$10 by the end of 2005, when V1647 Ori had dramatically faded, and \citet{asp08} reported $A_{V}$ was as high as 19 by early 2007.  Clearly, the extinction, as measured via optical and near-infrared measurements, changed first from quiescence to outburst and then from outburst back to quiescence.

These apparently discordant results have a straightforward and consistent explanation in the context of an accretion episode.  In X-rays, we are essentially detecting V1647 Ori along a direct line of sight to the stellar photosphere. Our results therefore indicate that the absorption along this direct line of sight --- which likely includes at least part of a thick circumstellar disk that is tilted about 30 degrees from edge-on \citep{aco07} --- does not change significantly as a function of time, despite the evident changes in X-ray luminosity.  The optical and near-infrared photons we observe, however, emerge from the near-photosphere environment of V1647 Ori along two paths.  One path, along our direct line of sight to the photosphere, produces heavily reddened and extincted light.  The second path takes photons nearly perpendicular to our line of sight, through an evacuated polar cavity, where they then scatter into our line of sight \citep{aco07}.  These photons are bluer and much less heavily extincted than the line-of-sight photons.  When V1647 Ori is in the quiescent state (1998, late 2005--2007), we see a faint, reddened, heavily extincted source because the contribution from scattered light is minimal. During the outburst state (2004--2005; 2008--2009), we see a brighter, bluer source because the contribution to the continuum of scattered photons is large.

\subsection{Possibility of a Second Plasma}

Though the plasma temperature of V1647 Ori strongly correlates with the X-ray luminosity and hardness ratio (Fig.~\ref{V1647OriLumHRTemp}), it is unclear whether we are observing a single-component plasma that increases or decreases in temperature and thus causes the observed changes in X-ray luminosity or if a second, lower-temperature plasma is also present and whose contribution to the total spectrum is overwhelmed by the hotter temperature plasma during outbursts.  Other sources, such as V710 Tau \citep{shu08} and EX Lupi \citep{gro10} have spectra that can be modeled as two-temperature plasmas, with one component fading as the star returns to quiescent levels.  \citet{gro05} was able to model the XMM-Newton observation of V1647 Ori in 2004 April with a single-temperature plasma but also found a better fit using a two-component model.  However, this observation of V1647 Ori was at least twice as long as most of the Chandra observations from 2004--2009.  In addition, XMM-Newton is more sensitive to lower-energy photons than CXO/ACIS.  Therefore, it may be the case that V1647 Ori does have two plasmas that contribute to the X-ray spectra but that ACIS was unable to consistently detect the lower-temperature component due to the inherit limitations of its design, the shorter observation times of the CXO observations, and the high degree of softer X-ray absorption by the intervening hydrogen column.

In order to test whether CXO would be able to detect two distinct plasma temperatures in 5 ks and 20 ks observations, we simulated a two-component plasma (using \emph{fakeit}) with a normalization ratio and plasma temperatures identical to those found by \citet{gro05} and suffering extinction by the same hydrogen column density.  The simulated spectra were then convolved with the responses of the front-illuminated ACIS-I3 and back-illuminated ACIS-S3 CCDs.  The simulations incorporated appropriately-scaled normalization parameters so that the resultant count rates were comparable to those of the 2005 and 2008--2009 CXO observations.  Each of the simulated spectra was first fit with a single-component model and then with a two-component model.  All of the 20 ks observations were found to be fit better using two-component models with each fit yielding values for plasma temperatures and normalization that were close to the original simulation parameters.  F-tests also suggested a slight improvement in the model fits if the 75\%/25\% normalization ratio between the low/high components, as found by \citet{gro05}, was forced.  When fit with a two-component model, the simulated 5 ks observation of the more sensitive ACIS-S3 CCD yielded a best fit (using the 75\%/25\% normalization ratio) that converged to a model with two nearly-identical temperature plasmas, i.e., to a single-temperature plasma, and thus does not lead to a better fit.  Thus, it appears that CXO should have the sensitivity to detect a two-component plasma such as that found by \citet{gro05} if the observations are long enough and the lower-temperature plasma flux is strong enough  compared to the higher-temperature plasma flux.  

The spectral models for each of the five Chandra observations obtained in 2008--2009 consistently converge to fits with a single-temperature plasma that, to within the errors, is $\sim$4 keV.  As seen with the correlation between X-ray luminosity and the mean hardness ratio, median photon energy, and mean count rate, we also find that when the X-ray luminosity increases or decreases, so does the plasma temperature. This single-component model is consistent with the single-component model of the 2005 April 11 XMM-Newton data \citep{gro05}.   

\subsection{Similarity of the 2003 and 2008 Eruptions}

Finally, the second (2008) eruption of V1647 Ori has very similar spectral characteristics to that of the first (2003--2005) eruption.  From Tables \ref{tab-1}, \ref{tab-3}, and \ref{tab-2}, the five 2008--2009 observations of V1647 Ori, which span a period of approximately seven months, show 

\begin{itemize}
\item mean hardness ratios that are consistently as hard (to within the errors) as the hardest value of any observation following the previous eruption; 
\item median photon energies at levels that are very similar to the greatest median photon energy of any observation from the previous eruption;
\item mean count rates that are 20--50 times higher than those observed during X-ray quiescence and that are typically 1--2 times the highest mean count rate of the 2003--2006 observations;
\item plasma emission measures that are usually 1--2 times greater than any modeled from the 2003-2005 observations; 
\item and X-ray luminosities that are consistently greater than 1.5 $\times$ 10$^{30}$ ergs s$^{-1}$, 1--2 times the X-ray luminosity of V1647 Ori during any of the 2003-2005 CXO observations.
\end{itemize}
 
Given that the spectral characteristics of V1647 Ori are so similar in observations in which the X-ray flux is elevated above the X-ray quiescent level, it appears that the same X-ray generation mechanism was at work during both eruptions and that the plasma characteristics were very similar during the two eruption epochs.  The derived X-ray luminosities suggest that the second eruption was more energetic than the first, and given that the modeled emission measures are greater for the observations following the second eruption, it is reasonable to conclude that we observed the same phenomenon in both eruptions but that a larger mass of X-ray emitting plasma was active during the outburst that began in 2008.

\section{Summary}\label{summary}

Our X-ray monitoring demonstrates that the two optical/near-IR outbursts undergone by the enigmatic V1647 Ori in less than a decade were accompanied by strong X-ray outbursts. During these two outbursts (2003--2005 and 2008--present), we see that the X-ray flux rose to peak luminosity over a span of a few weeks and then remained elevated for approximately two years during the first eruption and for at least one year during the second eruption.  Given that there is very strong evidence that the outbursts observed in the optical and near-infrared regimes are driven by accretion, we conclude that the correlated outbursts in X-rays are also driven by accretion.  

We find that all of the CXO spectra of V1647 Ori are best modeled with a single, moderate-temperature (2--6 keV) plasma.  In almost all cases, the plasma temperature that emerges from models of the CXO spectra is too high to be generated via accretion hotspots on the stellar photosphere but is reasonable for a plasma generated via magnetic reconnection events. However, the X-ray-emitting plasma could also be located in a strongly enhanced stellar corona, or at the inner edge of the circumstellar disk.  Given that accretion is ongoing, lower-temperature plasma generated by shocks at the accretion footprint is very likely present; however, during these CXO observations, any such soft component contributed much less flux than the moderate-temperature plasma and so usually did not leave a distinct signature in the X-ray spectra.  With the elevated hard X-ray flux lasting the duration of the 2008--2009 epochs, we conclude that the X-ray flux is not the result of typical coronal flares generated via reconnection events. We believe that since the optical/near-infrared flux remains elevated throughout this observation epoch, we are instead observing the X-rays generated from reconnection events in the accretion stream, with the softer X-ray flux possibly being generated by accretion hotspots at the stellar photosphere.  

We find no significant change in X-ray absorbing column, indicating that varying optical/IR color measurements, which have previously been interpreted as evidence for variable reddening toward V1647 Ori, may instead be indicative of varying contributions from scattered vs. direct photospheric emission from the YSO.  Two of the spectra obtained during the most recent (2008) outburst appear to show the 6.4 keV neutral iron feature indicating fluorescence from cold (presumably circumstellar disk) gas surrounding V1647 Ori.  

With V1647 Ori being observed intensely during both outbursts at X-ray, optical, and infrared wavelengths, this objects stands as one of the best characterized systems that exhibits such a close correspondence between X-ray output and accretion rate.  As a result of intense monitoring at X-ray, optical, and infrared wavelengths during two successive accretion-driven outbursts, V1647 Ori stands as the best characterized YSO in terms of the correspondence between X-ray output and accretion rate. We have shown, furthermore, that this YSO exhibited strikingly similar X-ray behavior and spectral properties during its recent accretion bursts. These results underscore the need for X-ray monitoring of additional eruptive YSOs, so as to evaluate whether the remarkable consistency of V1647 Ori is the exception or the norm.

\acknowledgements
We thank Nuria Calvet for providing early access to the data in CXO ObsIDs 10763 and 8585. This research was supported via awards numbers GO8-9016X and GO9-0006X to Vanderbilt University issued by the Chandra X-ray Observatory Center, which is operated by the Smithsonian Astrophysical Observatory for and on behalf of NASA under contract NAS8-03060.

{\it Facilities:} \facility{CXO (ACIS)}

\clearpage

\newpage
\begin{deluxetable}{llcccccccc}
\setlength{\tabcolsep}{0.04in} 
\tabletypesize{\scriptsize}
\tablewidth{0pt}
\tablecaption{\label{tab-1}2002--2009 ACIS Observations.} 
\tablehead{\colhead{ObsID} & \colhead{Observation} & \colhead{JD} & \colhead{ACIS} & \colhead{Exposure} & \colhead{Net Counts} & \colhead{Mean Count} & \colhead{Median Photon} & \colhead{Mean Hardness\tablenotemark{a}} & \colhead{Hardness Ratio\tablenotemark{b}} \\
\colhead{ } & \colhead{Date} & \colhead{ } & \colhead{Chip} & \colhead{(ks)} & \colhead{(0.5--8.0 keV)} & \colhead{Rate (ks$^{-1}$)} & \colhead{Energy (keV)} & \colhead{Ratio} & \colhead{of Total Counts} }
\startdata
2539\tablenotemark{c} & 2002 Nov 14 & 2452593 & S2 & 62.8 & 17 & 0.3 $\pm$ 0.1 & 2.0 $\pm$ 0.3 & 			$-$0.34 $\pm$ 0.31  & $-$0.28 $\pm$ 0.28\\
5307\tablenotemark{d} & 2004 Mar 07 & 2453072 & S3 & 5.5 & 64 & 11.7 $\pm$ 1.5 & 3.6 $\pm$ 0.3 & 			0.46 $\pm$ 0.15 & 0.45 $\pm$ 0.14\\
5308\tablenotemark{d} & 2004 Mar 22 & 2453087 & S3 & 4.9 & 12 & 2.5 $\pm$ 0.8 & 2.0 $\pm$ 0.5  & 			$-$0.45 $\pm$ 0.33 & $-$0.46 $\pm$ 0.33 \\
5382 & 2005 Apr 11 & 2453472 & I3 & 18.2 & 86 & 4.8 $\pm$ 0.5 & 3.5 $\pm$ 0.1  & 						0.51 $\pm$ 0.19 & 0.46 $\pm$ 0.12\\
5383 & 2005 Aug 27 & 2453610 & I3 & 19.9 & 18 & 0.9 $\pm$ 0.2 & 3.0 $\pm$ 0.8  & 						0.01 $\pm$ 0.29 & 0.34 $\pm$ 0.25\\
5384 & 2005 Dec 09 & 2453714 & I3 & 19.7 & 2& 0.1 $\pm$ 0.1  & 2.2 $\pm$ 1.1  & 						0.0 $\pm$ 0.20 & $-$0.07 $\pm$ 0.92 \\
6413 & 2005 Dec 14 & 2453719 & I3 & 18.1 & 4& 0.2 $\pm$ 0.1 & 1.7 $\pm$ 0.3  &							$-$0.36 $\pm$ 0.26 & $-$1.0 $\pm$ 0.75\\
6414 & 2006 May 01 & 2453857 & I3 & 21.6 & 3 & 0.1 $\pm$ 0.1 & 1.3 $\pm$ 0.3  & 						0.09 $\pm$ 0.23 & 0.07 $\pm$ 0.92 \\
6415 & 2006 Aug 07 & 2453955 & I3 & 20.5 & 4 & 0.2 $\pm$ 0.1  & 2.4 $\pm$ 0.4  & 						$-$0.27 $\pm$ 0.21 & $-$1.0 $\pm$ 0.73 \\
\hline
9915 & 2008 Sep 18 & 2454728 & I3 & 19.9 & 466 & 23.6 $\pm$ 1.1 & 3.5 $\pm$ 0.1 & 						0.36 $\pm$ 0.07 & 0.37 $\pm$ 0.05 \\
10763\tablenotemark{c} & 2008 Nov 27 & 2454798 & I2 & 19.7 & 217& 11.0 $\pm$ 0.7 & 3.9 $\pm$ 0.1  & 		0.65 $\pm$ 0.09  & 0.64 $\pm$ 0.08\\
8585\tablenotemark{c} & 2008 Nov 28 & 2454799 & I2 & 28.5 & 160 & 5.8 $\pm$ 0.4 & 3.4 $\pm$ 0.1   & 		0.40 $\pm$ 0.10  & 0.43 $\pm$ 0.09 \\
9916 & 2009 Jan 23 & 2454855 & I3 & 18.4 & 245 & 13.6 $\pm$ 0.9 & 3.7 $\pm$ 0.2  & 						0.41 $\pm$ 0.08  & 0.43 $\pm$ 0.07\\
9917 & 2009 Apr 21 & 2454943 & I3 & 29.8 & 260 & 8.8 $\pm$ 0.5 & 3.5 $\pm$ 0.2  & 						0.28 $\pm$ 0.08  & 0.37 $\pm$ 0.07\\
\enddata
\tablenotetext{a}{Average of hardness ratios computed from 2 ks light curve data bins (ObsID 2539 uses 10 ks light curve data bins).}
\tablenotetext{b}{Hardness ratio computed using the total numbers of hard and soft X-ray photons from the entire observation.}
\tablenotetext{c}{V1647 Ori was not the target of the observation, and given values have been adjusted to compensate for the 4$\arcmin$ off-axis position of V1647 Ori.  Exposure maps indicated net counts and mean count rates for ObsID 2539 should be increased by 8\% while those for ObsIDs 10763 and 8585 required a 3\% increase.}
\tablenotetext{d}{The back-illuminated ACIS-S3 CCD is more sensitive to X-rays from plasma in the temperature regime characteristic of V1647 Ori than are the front-illuminated ACIS CCDs.  Values displayed for net counts and mean count rates, and their associated errors, for ObsIDs 5307 and 5308 have been adjusted downward by 10\%, based on spectral simulations in order to make net counts and mean count rates directly comparable between all chips.}
\tablecomments{All errors are 1$\sigma$.  The net counts for each observation are the total number of counts within the 0.5--8.0 keV range.  Median photon energy uncertainties were calculated via the half-sample method used in \citet{kas06}.  Mean count rates were determined by dividing the net counts by exposure times.  Uncertainties for mean count rates and hardness ratios of total counts follow Poisson statistics.  The uncertainty for the hardness ratio of the total counts for ObsID 5384 could not be calculated due to the even distribution of the very low number of counts}

\end{deluxetable}

\newpage
\begin{deluxetable}{ccccc}
\setlength{\tabcolsep}{0.1in} 
\tabletypesize{\scriptsize}
\tablewidth{0pt}
\tablecaption{\label{tab-fibi}Comparison of Simulated Front- and Back-Illuminated Chip Exposures} 
\tablehead{\colhead{Plasma} & \colhead{S3/I3} & \colhead{S3} & \colhead{I3} & \colhead{HR} \\
\colhead{Temperature (kev)} & \colhead{Count Ratio} & \colhead{HR} & \colhead{HR} & \colhead{Difference}}
\startdata
2 & 1.21 & 0.02 & 0.12 & 0.10 \\
5 & 1.14 & 0.30 & 0.38 & 0.08 \\
7 & 1.11 & 0.40 & 0.48 & 0.08 \\

\enddata
\tablecomments{Net count and hardness ratio (of total counts) comparisons for simulated spectra convolved with instrument responses from ObsIDs 5307 and 9915, which used the back-illuminated S3 CCD and front-illuminated I3 CCD, respectively.  Each of the 1-ks spectra were simulated using the \emph{fakeit} command in XSPEC with the intervening hydrogen column density and chemical abundance fixed at 4.0 $\times$ 10$^{22}$ cm$^{-2}$ and 0.8 solar, respectively.  The front-illuminated chip detects roughly 85\% of the net counts of the back-illuminated chip, with a greater drop in sensitivity for soft versus hard X-rays.  }

\end{deluxetable}

\newpage
\begin{deluxetable}{lcccccccc}
\setlength{\tabcolsep}{0.03in} 
\tabletypesize{\scriptsize}
\tablewidth{0pt}
\tablecaption{\label{tab-3}Model Fits for 2008--2009 Chandra Observations.}
\tablehead{\colhead{ObsID} & \colhead{Observation} & \colhead{Reduced} & \colhead{Degrees of } & \colhead{N$_{H}$} & \colhead{kT$_{X}$}   & \colhead{EM} & \colhead{Observed F$_{X}$} & \colhead{Observed L$_{X}$}  \\

\colhead{ } & \colhead{Date } & \colhead{$\chi$$^{2}$ } & \colhead{Freedom } & \colhead{($\times$ 10$^{22}$ cm$^{-2}$)} & \colhead{(keV)} & \colhead{($\times$ 10$^{53}$ cm$^{-3}$)} & \colhead{ ($\times$ 10$^{-13}$ ergs cm$^{-2}$ s$^{-1}$)} & \colhead{($\times$ 10$^{30}$ ergs s$^{-1}$) } }

\startdata
9915 & 2008 Sep 18  & 0.85 & 77 & 4.1${+0.9\atop-0.7}$ & 3.7${+2.3\atop-1.0}$ 	 & 15.0${+6.3\atop-4.4}$ & 4.2${+0.3\atop-1.1}$ & 8.1${+0.5\atop-2.1}$ \\
10763 & 2008 Nov 27 &  1.24 & 39 & 4.1 (Fixed) & 4.0 (Fixed)  				 & 6.1${+0.8\atop-0.8}$ & 1.8${+0.3\atop-0.3}$ & 3.5${+0.5\atop-0.5}$ \\	
8585 & 2008 Nov 28  & 0.71 & 26 &  5.9${+2.4\atop-1.4}$ & 2.0${+1.0\atop-0.6}$ 	 & 8.2${+9.5\atop-3.7}$ & 1.0${+0.1\atop-0.7}$ & 1.8${+0.3\atop-1.3}$ \\
9916 & 2009 Jan 23  &   0.79 & 42 & 4.1${+1.1\atop-0.8}$ & 6.0${+7.4\atop-2.6}$ 	& 6.8${+3.5\atop-1.6}$ & 2.6${+0.2\atop-1.4}$ & 5.0${+0.3\atop-2.7}$ \\
9917 & 2009 Apr 21  &  0.88 & 44 & 4.1${+1.3\atop-0.9}$ & 3.4${+2.1\atop-1.1}$ 	 & 5.9${+4.1\atop-2.0}$ & 1.5${+0.2\atop-0.7}$ & 2.9${+0.3\atop-1.3}$ \\
 
 \enddata
\tablecomments{Uncertainties given for hydrogen column density, plasma temperature, and chemical abundance correspond to the 90$\%$ confidence intervals, whereas the observed X-ray luminosities are given with their corresponding 68\% confidence intervals.  Chemical abundance was fixed at 0.8 solar.  For ObsID 10763, in order to constrain the observed X-ray flux and luminosity, the hydrogen column density and plasma temperature were fixed at the given values.  Emission measures and luminosities assume a distance of 400 pc to V1647 Ori.}
\end{deluxetable}

\newpage
\begin{deluxetable}{lcccccccc}
\setlength{\tabcolsep}{0.05in} 
\tabletypesize{\scriptsize}
\tablewidth{0pt}
\tablecaption{\label{tab-2}Model Fits for 2002--2006 Chandra Observations.}
\tablehead{\colhead{ObsID} & \colhead{Observation} & \colhead{Reduced} & \colhead{Degrees of }  & \colhead{kT$_{X}$}  & \colhead{EM} & \colhead{Observed F$_{X}$} & \colhead{Observed L$_{X}$}  \\
\colhead{ } & \colhead{Date } & \colhead{Statistic } & \colhead{Freedom }  & \colhead{(keV)} & \colhead{($\times$ 10$^{53}$ cm$^{-3}$)} & \colhead{ ($\times$ 10$^{-13}$ ergs cm$^{-2}$ s$^{-1}$)} & \colhead{($\times$ 10$^{30}$ ergs s$^{-1}$) }  }
\startdata
2539	&	2002	Nov	14	&	1.37	&	13		&	0.9${+1.1\atop-0.5}$	&	0.6${+1.9\atop-0.6}$	&	0.03${+0.01\atop-0.02}$	&	0.05${+0.01\atop-0.04}$\\
5307	&	2004	Mar	7	&	1.19\tablenotemark{a}	&	10		&	4.0 (F)	&	5.4 $\pm$ 1.3	&	1.6 $\pm$ 0.5	&	3.1 $\pm$ 0.9\\
5308	&	2004	Mar	22	&	1.45	&	10		&	1.1${+1.0\atop-0.5}$	&	3.9${+6.9\atop-3.9}$	&	0.2 $\pm$ 0.1\tablenotemark{b}	&	0.4 $\pm$ 0.1\\
5382	&	2005	Apr	11	&	0.98\tablenotemark{a}	&	14	&		3.4${+3.0\atop-1.1}$	&	3.1${+1.2\atop-1.0}$	&	0.8${+0.1\atop-0.3}$	&	1.5${+0.2\atop-0.6}$\\
5383	&	2005	Aug	27	&	1.99	&	16		&	3.2${+10.1\atop-1.7}$	&	0.8${+0.8\atop-0.4}$	&	0.2${+0.1\atop-0.2}$	&	0.4${+0.1\atop-0.3}$\\
5384	&	2005	Dec	9	&	1.67	&	1		&	0.86 (F)	&	0.2${+0.6\atop-0.2}$	&	0.01 $\pm$ 0.01\tablenotemark{b}	&	0.01 $\pm$ 0.01\\
6413	&	2005	Dec	14	&	0.87	&	3		&	0.86 (F)	&	0.9${+0.8\atop-0.5}$	&	0.02 $\pm$ 0.01\tablenotemark{b}	&	0.05 $\pm$ 0.03\\
6414	&	2006	May	1	&	2.83	&	2		&	0.86 (F)	&	0.3${+0.6\atop-0.3}$	&	0.01 $\pm$ 0.01\tablenotemark{b}	&	0.02 $\pm$ 0.02\\
6415	&	2006	Aug	7	&	1.45	&	3		&	0.86 (F)	&	0.9${+0.6\atop-0.3}$	&	0.01 $\pm$ 0.01\tablenotemark{b}	&	0.01 $\pm$ 0.01\\
\enddata

\tablenotetext{a}{Value is the reduced-$\chi$$^{2}$ value.}
\tablenotetext{b}{Observed X-ray flux errors and corresponding X-ray luminosity errors were derived by multiplying the mean count rate errors from Table \ref{tab-1} by an energy conversion factor (ECF).}
\tablecomments{Uncertainties given for hydrogen column density and plasma temperature correspond to the 90$\%$ confidence intervals, whereas the observed X-ray luminosities are given with their corresponding 68\% confidence intervals.  All models use a fixed hydrogen column density of N$_{H}$ = 4.1 $\times$ 10$^{22}$ cm$^{-2}$ and a chemical abundance fixed at 0.8 solar.  For ObsID 5307, the plasma temperature was fixed (``F") at a typical post-outburst plasma temperature of kT$_{X}$ = 4.0 keV in order to constrain the remaining model parameters.  For ObsIDs 5384, 6413, 6414, and 6415, the very low number of counts did not permit spectral fitting with freely-varying plasma temperature, so the plasma temperature was fixed at a pre-outburst value as discussed in \citet{kas04}.  Observed fluxes and emission measures were derived from the resulting model fits, an energy conversion factor (ECF) was calculated for each observation, and the mean count rate uncertainties were multiplied with the ECF to better constrain the source fluxes displayed.  All models, unless otherwise noted, use the Cash statistic to determine goodness of fit, and the reduced statistic is the Cash-statistic or $\chi$$^{2}$ value divided by the number of degrees of freedom.  Emission measures and observed luminosities assume a distance of 400 pc to V1647 Ori.}
\end{deluxetable}

\newpage

\begin{figure}
\centering
\includegraphics[scale=0.5, angle=270, bb = 149 46 600 793]{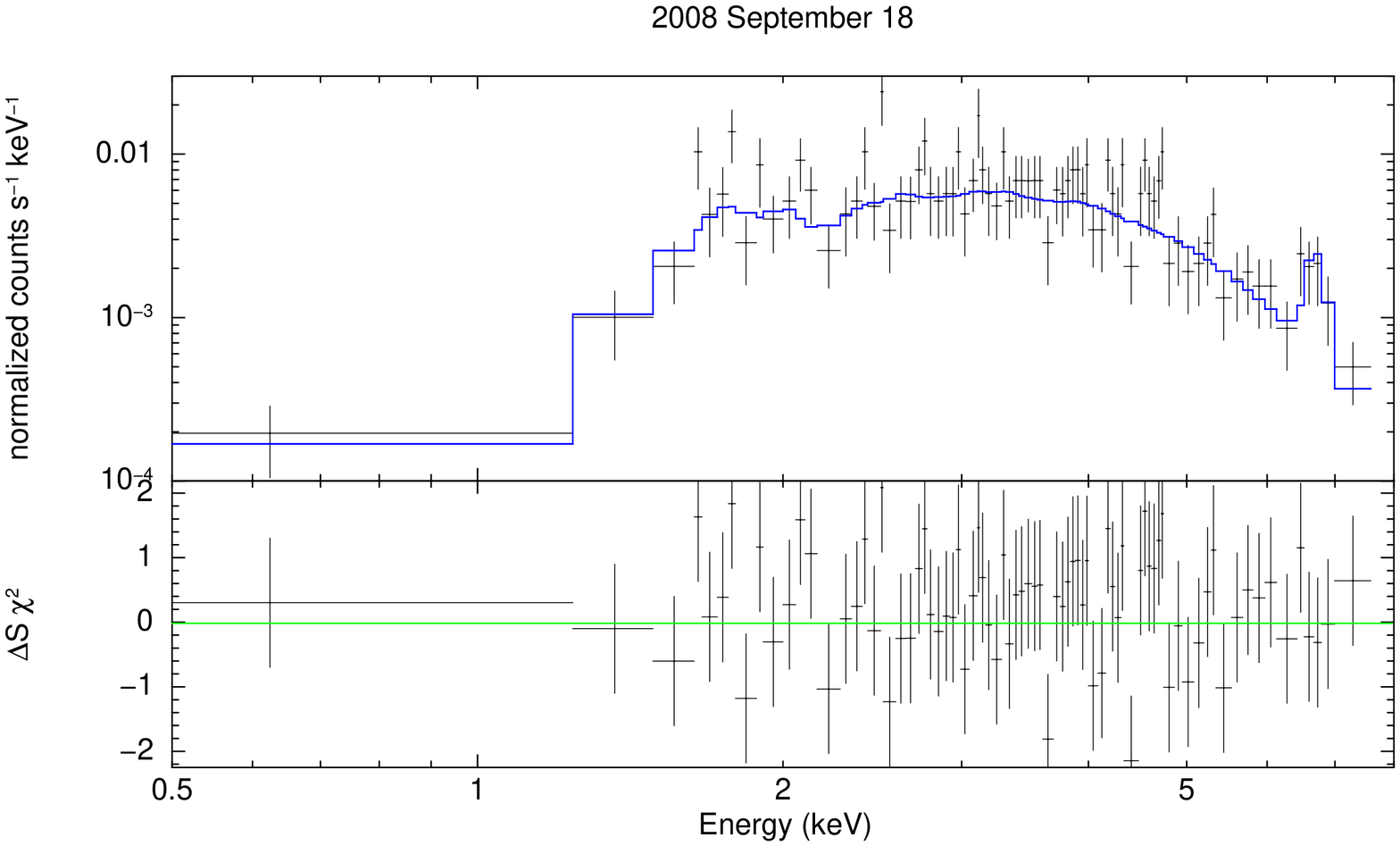}\\  
\includegraphics[scale=0.5, angle=270, bb = 149 46 600 793]{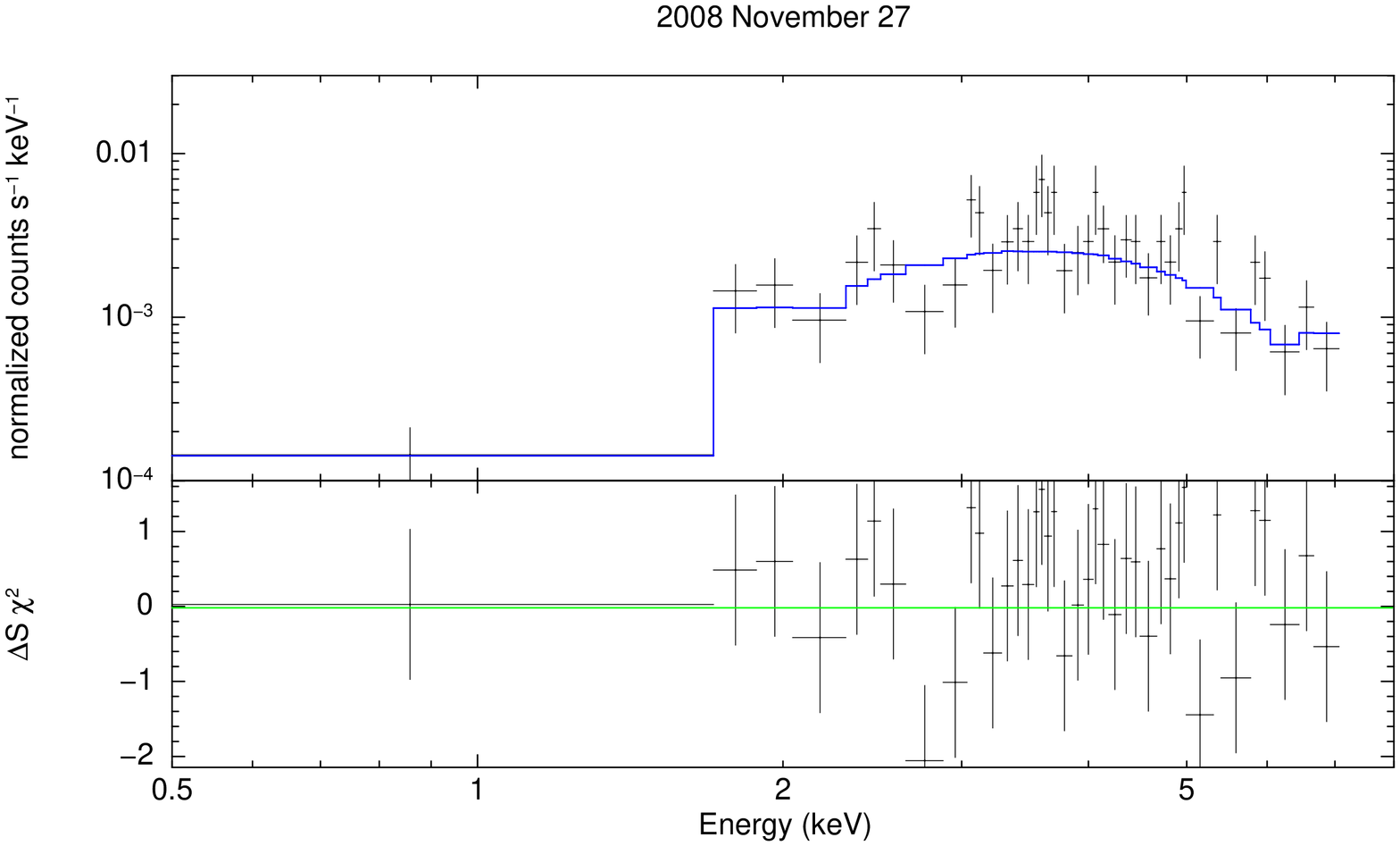}\\ 
\includegraphics[scale=0.5, angle=270, bb = 149 46 600 793]{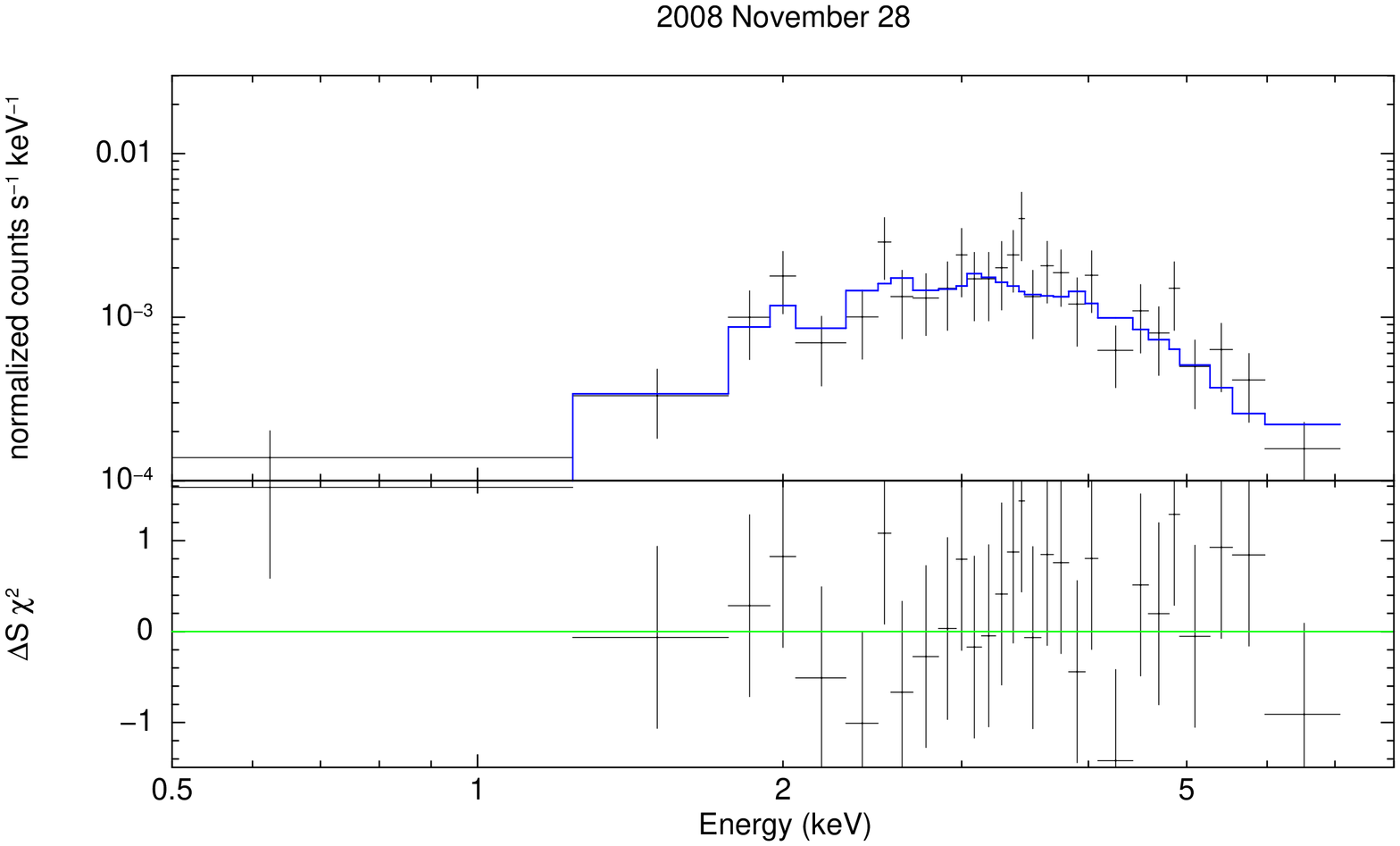}\\ 
\caption{}
\end{figure}

\begin{figure}
\ContinuedFloat
\centering
\includegraphics[scale=0.5, angle=270, bb = 149 46 600 793]{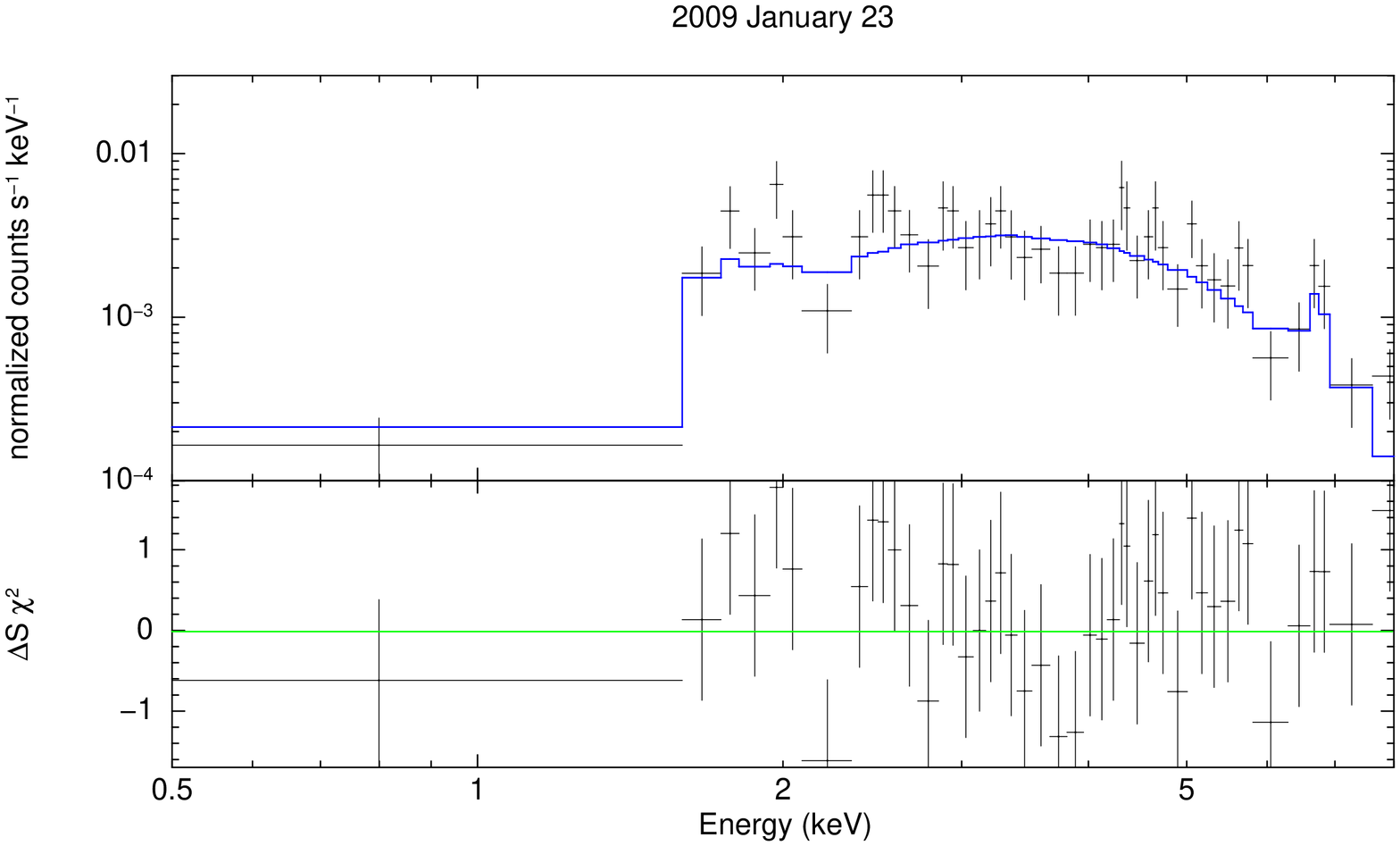}\\ 
\includegraphics[scale=0.5, angle=270, bb = 149 46 600 793]{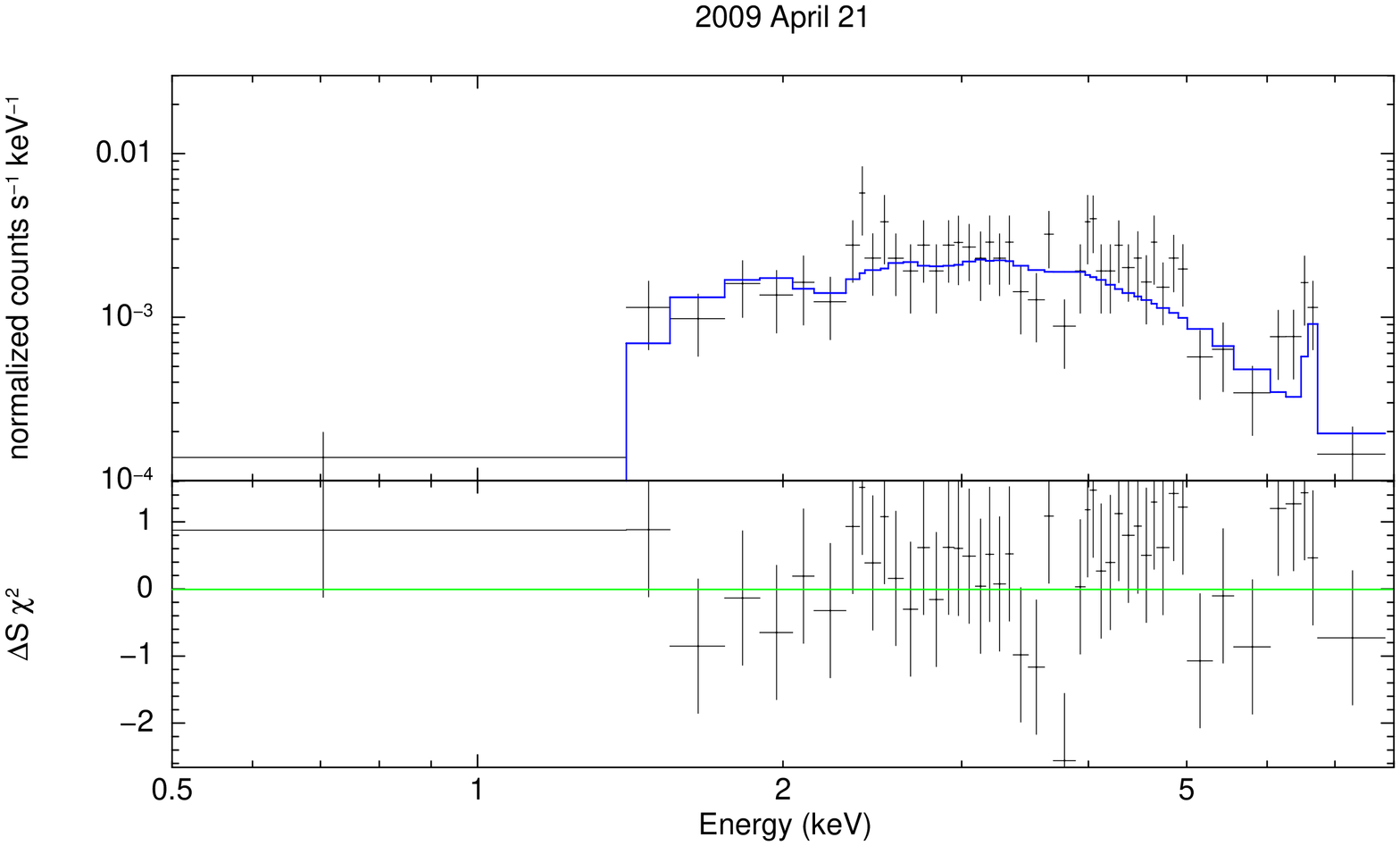}\\ 
\caption{Best-fit XSPEC models of spectra obtained from 2008 September to 2009 April, with observations in chronological order from top to bottom.  The top frame in each panel displays the data (binned to five-count-minimum bins) in black overlaid with the model in blue.  
\label{bestfive}}
\end{figure}

\newpage

\end{document}